\documentclass[twocolumn]{aastex631}
\usepackage{mathtools}

\accepted{February 25, 2023}

\shorttitle{NLTT5306}
\shortauthors{Amaro, Apai et al.}

\graphicspath{{./}{figures/}}

\begin{document}

\title{Hotter than Expected: HST/WFC3 Phase-resolved Spectroscopy of \\a Rare Irradiated Brown Dwarf with Strong Internal Heat Flux}

\correspondingauthor{Rachael Amaro}
\email{rcamaro@email.arizona.edu}

\author[0000-0002-1546-9763]{Rachael C. Amaro}
\altaffiliation{National Science Foundation Graduate Research Fellow}
\affiliation{Department of Astronomy and Steward Observatory, The University of Arizona, 933 North Cherry Avenue, Tucson, AZ 85721, USA}

\author[0000-0003-3714-5855]{D\'aniel Apai}
\affiliation{Lunar and Planetary Laboratory, The University of Arizona, 1640 E. University Blvd, Tucson, AZ 85721, USA}
\affiliation{Department of Astronomy and Steward Observatory, The University of Arizona, 933 North Cherry Avenue, Tucson, AZ 85721, USA}

\author[0000-0003-2969-6040]{Yifan Zhou}
\altaffiliation{51 Peg b Fellow}
\affiliation{Department of Astronomy, The University of Texas at Austin, 2515 Speedway, Austin, TX 78712, USA}

\author[0000-0003-1487-6452]{Ben W. P. Lew}
\affiliation{Bay Area Environmental Research Institute and NASA Ames Research Center, Moffett Field, CA 94035, USA}

\author[0000-0003-2478-0120]{Sarah L. Casewell}
\altaffiliation{STFC Ernest Rutherford Fellow}
\affiliation{School of Physics and Astronomy, University of Leicester, Leicester LE1 7RH, UK}

\author[0000-0002-4321-4581]{L. Mayorga}
\affiliation{The Johns Hopkins University Applied Physics Laboratory, 11100 Johns Hopkins Rd, Laurel, MD, 20723, USA}

\author[0000-0002-5251-2943]{Mark S. Marley}
\affiliation{Lunar and Planetary Laboratory, The University of Arizona, 1640 E. University Blvd, Tucson, AZ 85721, USA}

\author[0000-0003-2278-6932]{Xianyu Tan}
\affiliation{Atmospheric, Oceanic and Planetary Physics, Department of Physics, University of Oxford, Parks Road, OX1 3PU, Oxford, UK}

\author[0000-0003-3667-8633]{Joshua D. Lothringer}
\affiliation{Department of Physics, Utah Valley University, 800 W. University Parkway, Orem, UT, 84058, USA}

\author[0000-0001-9521-6258]{Vivien Parmentier}
\affiliation{Atmospheric, Oceanic and Planetary Physics, Department of Physics, University of Oxford, Parks Road, OX1 3PU, Oxford, UK}
\affiliation{Universit\'e C\^ote d'Azur, Observatoire de la C\^ote d'Azur, CNRS, Laboratoire Lagrange, France}

\author[0000-0002-7129-3002]{Travis Barman}
\affiliation{Lunar and Planetary Laboratory, The University of Arizona, 1640 E. University Blvd, Tucson, AZ 85721, USA}

\begin{abstract}
With infrared flux contrasts larger than typically seen in hot Jupiter, tidally-locked white dwarf-brown dwarf binaries offer a superior opportunity to investigate atmospheric processes in irradiated atmospheres. NLTT5306 is such a system, with a $M_{\rm{BD}}$=52$\pm$3 M$_{\rm{Jup}}$ brown dwarf orbiting a $T_{\rm{eff}}$=7756$\pm$35 K white dwarf with an ultra-short period of $\sim$102 min. We present \textit{HST}/WFC3 spectroscopic phase curves of NLTT5306, consisting of 47 spectra from 1.1 to 1.7 $\mu$m with an average S/N$\sim$65 per wavelength. We extracted the phase-resolved spectra of the brown dwarf NLTT5306B, finding a small $<$100~K day/night temperature difference ($\sim$5\% of the average day-side temperature). Our best-fit model phase curves revealed a complex wavelength-dependence on amplitudes and relative phase offsets, suggesting longitudinal-vertical atmospheric structure. The night-side spectrum was well-fit by a cloudy, non-irradiated atmospheric model while the day-side was best-matched by a cloudy, weakly irradiated model. Additionally, we created a simple radiative energy redistribution model of the atmosphere and found evidence for efficient day-to-night heat redistribution and a moderately high Bond albedo. We also discovered an internal heat flux much higher than expected given the published system age, leading to an age reassessment that resulted in NLTT5306B most likely being much younger. We find that NLTT5306B is the only known significantly irradiated brown dwarf where the global temperature structure is not dominated by external irradiation, but rather its own internal heat. Our study provides an essential insight into the drivers of global circulation and day-to-night heat transport as a function of irradiation, rotation rate, and internal heat.
\end{abstract}

\keywords{}

\section{Introduction} \label{sec:intro}


Spectroscopic phase curve observations of hot Jupiters \citep[e.g.,][]{Knutson12, Cowan12, Stevenson14, Arcangeli19, Mikal-Evans23} have revealed several hundred Kelvin temperature contrasts between their day and night hemispheres. These temperature contrasts drive a wealth of three-dimensional atmospheric dynamics, such as waves, jets, and turbulence, which have an impact on the three-dimensional atmospheric structure and temperature distribution \citep[see][for a review]{Showman20}.

Day-to-night heat transport on hot Jupiters is generally dominated by equatorial jets \citep[][]{ShowmanGuillot02}. A number of recent studies have been dedicated to understanding how this atmospheric heat transport is affected by various parameters including rotation rate, irradiation strength, and atmospheric composition \citep[][]{TanShowman20_rotationWDBDs, parmentier2021, GaoPowell21, Komacek2022, lian2022}.
Tackling these questions requires high S/N observations capable of measuring heat transport and a sample of objects that continuously span the parameters mentioned above. High S/N observations can be challenging to obtain for hot Jupiters due to the low relative brightness of their host stars. However, there exists another class of objects that experiences levels of irradiation similar to hot Jupiters -- white dwarf-brown dwarf (WD+BD) binaries with ultra-short orbital periods \citep[][]{Casewell12, Beuermann13, Parsons17, Casewell20_WD1032, Casewell20}. The WD hosts in these binaries not only emit at shorter wavelengths than their main sequence counterparts in hot Jupiter systems, but are also much smaller in size. So, while hot Jupiters suffer from small infrared flux contrasts with their host stars, irradiated brown dwarfs typically exhibit infrared flux contrasts that are at least two orders of magnitude better. This makes WD+BD binaries perfect proxies to hot Jupiters.

WD+BD pairs with ultra-short periods are thought to be the result of common envelope stellar evolution, leaving behind tight binaries that will eventually become cataclysmic variables \citep[][]{Politano2004}. The extremely short periods of these systems ($\sim$1$-$2 hours) allows observations to capture more orbits in a shorter amount of time, leading to more robust analyses and results. In atmospheric terms, irradiated brown dwarfs act as a bridge between non-irradiated brown dwarfs and irradiated hot Jupiters. For most of the ultra-short period WD+BD systems that have been studied so far, day-side emission is largely dominated by irradiation over internal heat, e.g., WD~0137B \citep[][]{Maxted06}, SDSS J1205-0242B \citep[][]{Parsons17}, and EPIC~2122B \citep[][]{Casewell18a}. Thus, observationally testing the interplay between interior heat flux and irradiation along with the impact it has on atmospheric dynamics and thermal structures has not yet been possible.

\begin{figure}
\begin{center}
\includegraphics[width=0.47\textwidth]{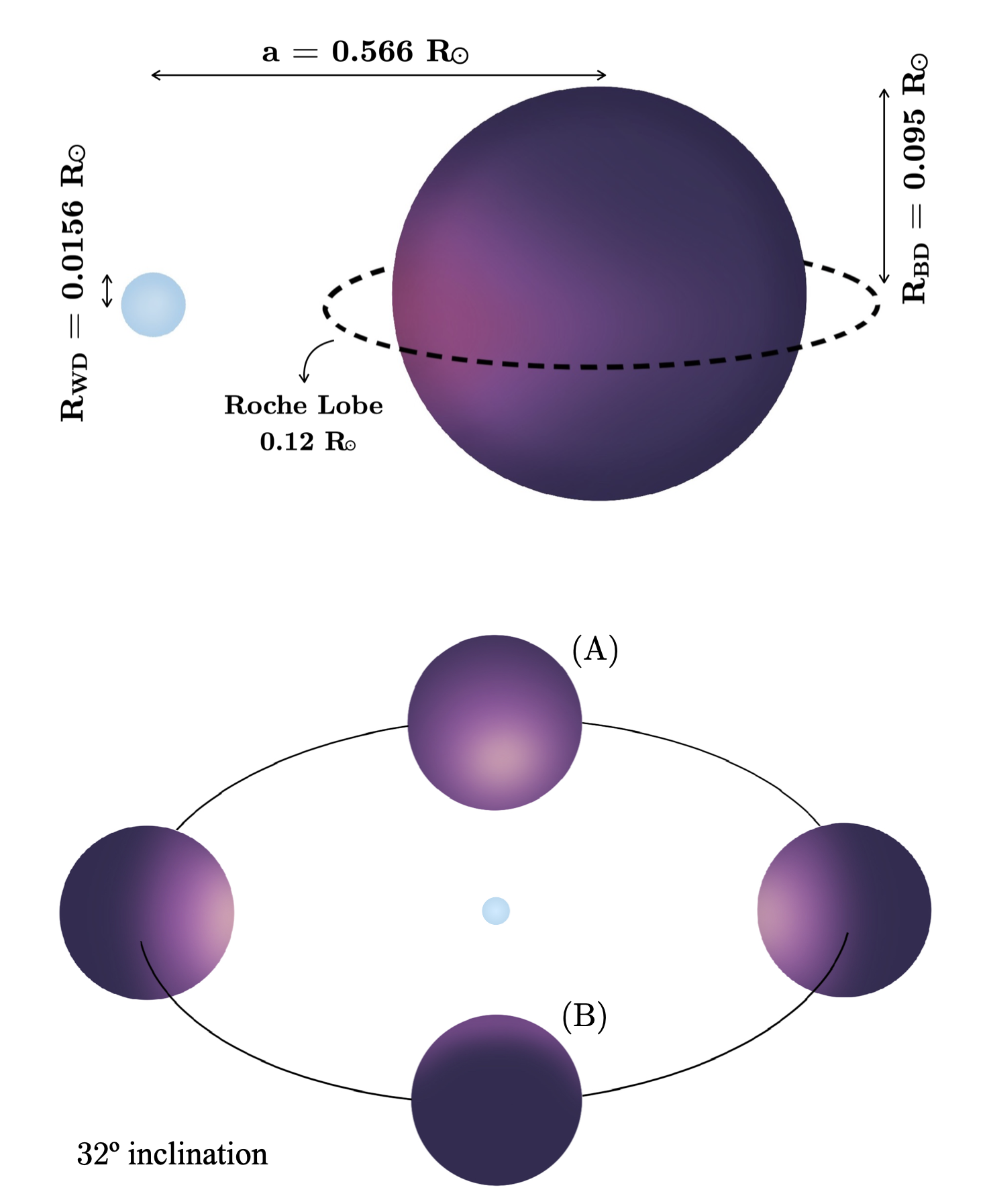}
\caption{Schematic of NLTT5306. \textbf{Top:} Near edge-on view of NLTT5306, showing relative sizes for the white dwarf, brown dwarf, and brown dwarf Roche Lobe radius. Separation between the white dwarf and brown dwarf are not to scale. Colors correspond to calculated temperatures from spectral fitting. \textbf{Bottom:} NLTT5306 viewed from a hypothesized 32$^{\circ}$ inclination. To an observer, the day-side of the brown dwarf would be seen when the brown dwarf is in the position labeled (A). Conversely, night would be when the brown dwarf is at the position labeled (B).}
\label{fig:schematic}
\end{center}
\end{figure}

\begin{deluxetable*}{llcl}
\tablecaption{Properties of the NLTT5306 binary system \label{tab:keyprops}}
\tablewidth{10pt}
\tablehead{
\colhead{Parameter} & \colhead{Units} & \colhead{Value} & \colhead{Reference}
}
\startdata
$a$ & Orbital Separation [R$_{\odot}$] & 0.566$\pm$0.005 & \citet{Steele13} \\
$i$ & Orbital Inclination [deg] & 78.0$^{+1.1}_{-7.6}$ & This work \\
$D$ & Distance [pc] & 76.96$^{+0.55}_{-0.53}$ & \citet{Casewell20} \\
$P$ & Orbital Period [mins] & 101.88$\pm$0.02 & \citet{Steele13} \\
$T_{\rm{eff,WD}}$ & Effective Temperature [K] & 7756$\pm$35 & \citet{Steele13} \\
$T_{\rm{eff,BD}}$ (Day) & Effective Temperature [K] & $\sim$1640 & This work \\
$T_{\rm{eff,BD}}$ (Night) & Effective Temperature [K] & $\sim$1500 & This work \\
Spectral Type (BD) &  & L5 & \citet{Casewell20} \\
log~$g$ (WD) & Surface Gravity [cm s$^{-2}$] & 7.68$\pm$0.08 & \citet{Steele13} \\
$M_{\rm{WD}}$ & Mass [M$_{\odot}$] & 0.44$\pm$0.04 & \citet{Steele13} \\
$M_{\rm{BD}}$ & Mass [M$_{\odot}$] & 0.05$\pm$0.003 & \citet{Steele13} \\
Cooling Age (WD) & [Myr] & 710$\pm$50 & \citet{Steele13} \\
$R_{\rm{WD}}$ & Radius [R$_{\odot}$] & 0.0156$\pm$0.0016 & \citet{Steele13} \\
$R_{\rm{BD}}$ & Radius [R$_{\odot}$] & 0.095$\pm$0.004 & \citet{Steele13} \\
$R_{\rm{RL,BD}}$ & Roche Lobe Radius [R$_{\odot}$] & 0.12$\pm$0.02 & \citet{Longstaff19} \\
\enddata
\end{deluxetable*}

\vspace{-0.9cm}
Atmospheric properties of highly irradiated brown dwarfs have been investigated through 3D circulation models and 1D radiative-convective models \citep[][]{TanShowman20_rotationWDBDs, Lothringer_Casewell20, Lee20}. \citet{TanShowman20_rotationWDBDs} and \citet{Lee20} both found that the fast rotation rates on irradiated brown dwarfs significantly shrink the width of the predicted equatorial jet and suppress the day-to-night heat transfer. They also found that the day-side hot spot is not subject to eastward-shifting, unlike in hot Jupiter atmospheres \citep[][]{Knutson07, Showman11, Heng15, Wong16}. \citet{Lothringer_Casewell20} created self-consistent model spectra of brown dwarf atmospheres, incorporating the effect of strong ultraviolet irradiation. They found that in brown dwarf atmospheres that are irradiation-dominated, molecules are efficiently dissociated by the intense irradiation, resulting in atmospheres that resemble ultra-hot Jupiters \citep[][]{Arcangeli18, Lothringer18, Parmentier18, Kitzmann18}, i.e., no molecular absorption on the day-side. So far, theoretical studies of irradiated brown dwarfs have only considered scenarios in which the brown dwarf receives strong external irradiation. Additionally, all significantly irradiated atmospheres studied to date via high-precision time-series spectroscopy have been irradiation-dominated: WASP-43b \citep[][]{Stevenson14}; WASP-18b \citep[][]{Arcangeli19}; WASP-12b \citep[][]{Arcangeli2021}; SDSS1411-B \citep[][]{Lew22}; WD~0137B and EPIC~2122B \citep[][]{Zhou22}; and WASP-103b \citep[][]{Kreidberg18}.

To further our understanding of irradiated substellar atmospheres and their dependencies on rotation rate, internal heat flux, and external irradiation strength, we present Hubble Space Telescope (\textit{HST}) Wide Field Camera 3 (WFC3) observations of the WD+BD binary system LSPM-J0135+1445 (hereafter NLTT5306). These observations, and those described in \citet{Lew22} and \citet{Zhou22}, are part of the \textit{HST} program GO-15947 (PI: Apai). NLTT5306 is a detached binary, estimated to be filling $\approx$80\% of its Roche Lobe radius (Figure~\ref{fig:schematic}). NLTT5306 also occupies a unique parameter space when compared to the other WD+BD binaries that have already been studied in that its irradiation level is at least an order of magnitude lower, yet its internal heat flux is still quite strong \citep[][]{Showman16}.

This paper is organized as follows: In Section~\ref{sec:bdwd_system} we introduce WD+BD system NLTT5306 and relevant findings from previous studies. We describe the details of our observations in Sections~\ref{sec:observations} and data reduction procedures in \ref{sec:dataredux}. In Section~\ref{sec:lcanalysis}, we present the broadband light curve along with synthetic phase curves and evidence of relative phase offsets; in Section \ref{sec:bdspec}, we extract the phase-resolved spectra of NLTT5306B and derive brightness temperatures from the day- and night-sides. We compare the day and night spectra to field brown dwarfs in Section~\ref{sec:compare_fieldBDs}. In Sections~\ref{sec:Compare_nonirr_models} and \ref{sec:Compare_irr_models}, we introduce atmospheric models of non-irradiated and irradiated brown dwarfs, respectively, and compare them to the day-and night-sides of NLTT5306B. In Section~\ref{sec:GCM}, we describe a cloud-free general circulation model and examine inconsistencies with observations. We discuss interpretations of our results in Section~\ref{sec:disc} and summarize our findings in Section~\ref{sec:conclusions}.

\begin{deluxetable*}{cllrc}
\tablecaption{Log of \textit{HST} WFC3 Observations for NLTT5306\label{tab:obslog}}
\tablewidth{0pt}
\tablehead{
\colhead{Orbit} & \colhead{Observation} & \colhead{Filter} & \colhead{Exp. Time} & \colhead{Exp. Start} \\
\colhead{} & \colhead{ Type} & \colhead{ID} & \colhead{[s]} & \colhead{[BJD$_{\rm{TDB}}$]}
}
\startdata
\hline
\multicolumn{5}{c}{Visit 1}\\
\hline
1 & Imaging & F132N & $2\times29.6=59.2$ & 2459096.4241 \\
1 & Spectroscopic & G141 & $8\times313.1=2504.8$ & 2459096.4254 \\
2 & Imaging & F132N & $2\times29.6=59.2$ & 2459096.4903 \\
2 & Spectroscopic & G141 & $8\times313.1=2504.8$ & 2459096.4917 \\
3 & Imaging & F132N & $2\times29.6=59.2$ & 2459096.5566 \\
3 & Spectroscopic & G141 & $8\times313.1=2504.8$ & 2459096.5579 \\
\hline
\multicolumn{5}{c}{Inter-visit gap: 797.4 min}\\
\hline
\multicolumn{5}{c}{Visit 2}\\
\hline
4 & Imaging & F132N & $2\times29.6=59.2$ & 2459097.1522 \\
4 & Spectroscopic & G141 & $8\times313.1=2504.8$ & 2459097.1536 \\
5 & Imaging & F132N & $2\times29.6=59.2$ & 2459097.2185 \\
5 & Spectroscopic & G141 & $8\times313.1=2504.8$ & 2459097.2199 \\
6 & Imaging & F132N & $2\times29.6=59.2$ & 2459097.2848 \\
6 & Spectroscopic & G141 & $8\times313.1=2504.8$ & 2459097.2861 \\
\enddata
\tablecomments{BJD$_{\rm{TDB}}$ here was calculated by converting from UT start date and time of observations (DATE-OBS and TIME-OBS in MJD in the \texttt{flt} fits headers) using an online applet developed by Jason Eastman \citep[]{Eastman10_UTC2BJD}.}
\end{deluxetable*}

\vspace{-0.9 cm}
\section{White Dwarf Brown Dwarf\\ Binary System NLTT5306} \label{sec:bdwd_system}
NLTT5306 was first suggested to be a binary by \citet{Girven11} and \citet{Steele11} who noted the WD exhibited a near-infrared (NIR) excess, which led to the discovery of the brown dwarf companion. Key physical properties of the system are reported in Table~\ref{tab:keyprops}.

\citet{Steele13} determined $T_{\rm{eff}}$ = 7756$\pm$35 K and log($g$) = 7.68$\pm$0.08 for the WD, making it the coolest known WD host in a post-common envelope binary \citep[Figure 5 in][]{Steele13}. They reported a mass of $M_{\rm{WD}}$ = 0.44$\pm$0.04 M$_{\odot}$ and a photometric distance of $D$ = 71$\pm$4 pc. Using radial velocity measurements, the best-fit period was $P$ = 101.88$\pm$0.02 min with a minimum BD mass of $M_{\rm{BD}}$ = 56$\pm$3 M$_{\rm{Jup}}$. A $T_{\rm{eff, BD}}$ = 1700 K suggested a spectral type of L4-L7, which gave a radius of $R_{\rm{BD}}$ = 0.95$\pm$0.04 R$_{\rm{Jup}}$. \citet{Longstaff19} refined the ephemeris and the period to $P$ = 0.07075025(2) days and further constrained the spectral type of the BD to be between L3-L5, consistent with \citet{Steele13}. More recently \citet{Casewell20} compared an IRTF SpeX spectrum, integrated over a large portion of the orbit, with spectra of standard L dwarf spectra, finding a best match of spectral type of L5. \citet{Casewell20} also matched observed magnitudes for the BD alone with those given for field brown dwarfs in \citet{DupuyLiu12}, determining a spectral type range of L3$-$L5 for NLTT5306B.

From the lack of any eclipse in the NLTT5306 system, \citet{Buzard22} constrained the inclination to be below $i$=78.7$\pm$0.4$^{\circ}$ (where 90$^{\circ}$ is edge-on). \citet{Buzard22} also fit models to 44 epochs of NIR spectra from Keck+NIRSPEC \citep[][]{McLean98_NIRSPEC, Martin18_NIRSPEC}, as well as two subsets that separated the day- and night-side of NLTT5306B. Their results showed a minimal difference between day- and night-sides for effective temperature. Interestingly, they also found evidence from one set of models to support low-gravity on the day-side alone, yet a different set of irradiated models found no significant difference in surface gravity between day and night.

Figure~\ref{fig:schematic} shows a schematic diagram of NLTT5306 with an almost edge-on configuration and a hypothesized 32$^{\circ}$ inclination, using parameters presented in Table~\ref{tab:keyprops} and temperatures calculated in this analysis. The diagram shows relative sizes for the white dwarf, brown dwarf, and brown dwarf Roche Lobe radius. However, orbital separation between the white dwarf and brown dwarf are not to scale. The color of the white dwarf came from \citet{HarreHeller21} and its published effective temperature of 7756~K. The color of the brown dwarf came from \citet{Cranmer21} and the estimated day- and night-side temperatures.

\begin{figure*}
\begin{center}
\includegraphics[width=0.87\textwidth]{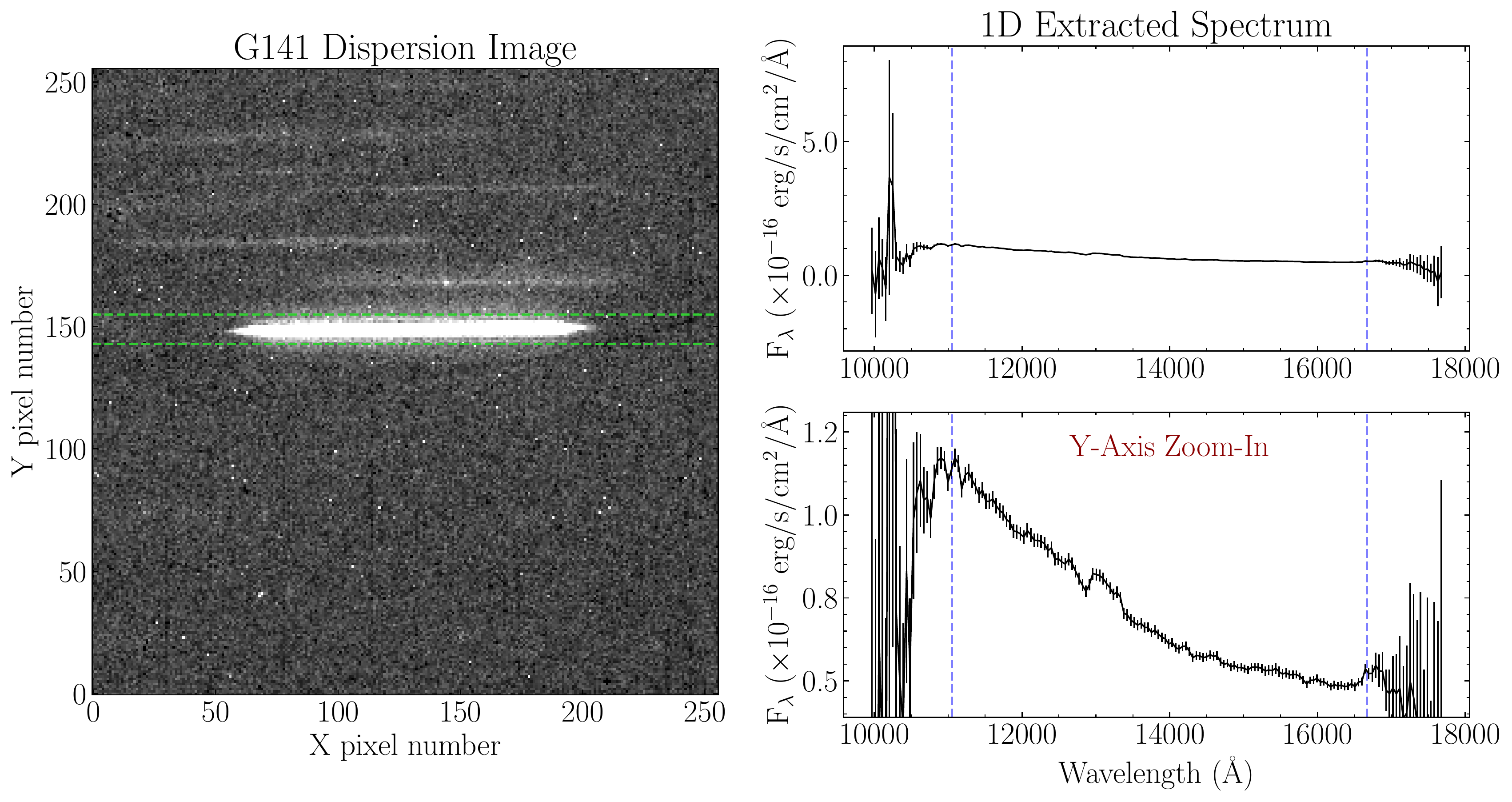}
\caption{\textbf{Left:} G141 grism observation of NLTT5306 system. The brightest horizontal line, around Y=150 and between the green dashed lines, is the grism spectrum of NLTT5306. For transparency, this image shows the data before cosmic ray removal and includes bad pixels that are corrected before \texttt{aXe} performs spectral extraction (see Section~\ref{sec:dataredux}). \textbf{Top Right:} An example of a 1D spectrum from \texttt{axecore} using a 4 pixel radius extraction window. Spectra inside the vertical, blue dashed lines denote the wavelengths used for science, approximately 1.1 to 1.67 $\mu$m. \textbf{Bottom Right:} Same as top left, except the y-axis has been zoomed in to highlight the shape of the science spectrum. The exceptionally high signal-to-noise spectra from \textit{HST} WFC3 allowed us study the system in remarkable detail.}
\label{fig:dispimage}
\end{center}
\end{figure*}

\section{Observations} \label{sec:observations}
As presented in Table~\ref{tab:obslog}, NLTT5306 was observed over 2 visits of 3 orbits each on September 3rd and 4th, 2020 with Hubble Space Telescope (\textit{HST}) using the WFC3/IR/G141 grism as part of \textit{HST} program GO-15947 (PI: Apai). Each visit lasted approximately 236 minutes with a $\sim$797 minute gap between the end of Visit~1 and the start of Visit~2. The observations were scheduled this way to provide near-complete orbital and rotational phase coverage for this system, since the orbital period of NLTT5306 was very close to that of \textit{HST}.

At the beginning of each orbit, two direct images were obtained for wavelength calibration using the F132N filter, a 256$\times$256 subarray setup, and the GRISM256 aperture. After direct images, each orbit acquired eight spectroscopic exposures in ``Staring mode'' of 313 s each, using the G141 grism, a 256$\times$256 subarray setup, and the GRISM256 aperture (left panel of Figure~\ref{fig:dispimage}). This observing sequence was successfully carried out for three other WD+BD systems as part of the same \textit{HST} program: \object{WD~0137-349}, \object{EPIC~212235321} \citep[]{Zhou22}, and SDSS~J141126.20$+$200911.1 \citep[]{Lew22}.

The third G141 observation in Orbit 6 was found to be strongly affected by a diagonal streak affecting approximately one-fourth of the image. The streak intersected with the target data and contaminated the extracted spectrum of NLTT5306. Given the 47 other high-quality frames, we excluded this frame from our subsequent analysis.

\section{Data Reduction} \label{sec:dataredux}

To start the spectral extraction process, we downloaded the \texttt{CalWFC3} (version 3.5.2) pipeline product \texttt{flt} files from the Barbara A. Mikulski Archive for Space Telescopes (MAST)\footnote{\url{https://archive.stsci.edu/index.html}}. We then reduced the data using an established pipeline, which combines the WFC3/IR spectroscopic software \texttt{aXe} \citep[]{Kummel09_aXe} and custom \texttt{Python} script. This pipeline has been successful in extracting time-resolved observations of brown dwarfs \citep[e.g.,][]{Buenzli12, Apai13} as well as time-resolved observations of WD+BD binaries \citep[e.g.,][]{Zhou22, Lew22}. 

We began by organizing the data into their respective orbits, so the G141 observations in a certain orbit can have highly precise wavelength calibration using its respective direct images. We then corrected bad pixels in the G141 data, identified via the Data Quality (DQ) extension, and then linearly interpolated with adjacent good pixels. The linear interpolation was performed in the x and y direction using the median value of the neighboring four pixels on each side (i.e., up to sixteen pixels). This method worked well everywhere except for pixels that are within four pixels of the subarray edge, however, this did not effect our target data.

The F132N images and G141 data were then embedded into larger full-frame-sized (1,014$\times$1,014) arrays to allow \texttt{aXe} to use the standard full-frame calibration images. We median combined the full-frame F132N direct images and performed precise source extraction \citep[]{Bertin96_sextractor}. The output catalog file was then used by \texttt{aXe} for wavelength calibration.

\begin{figure}
\begin{center}
\includegraphics[width=0.47\textwidth]{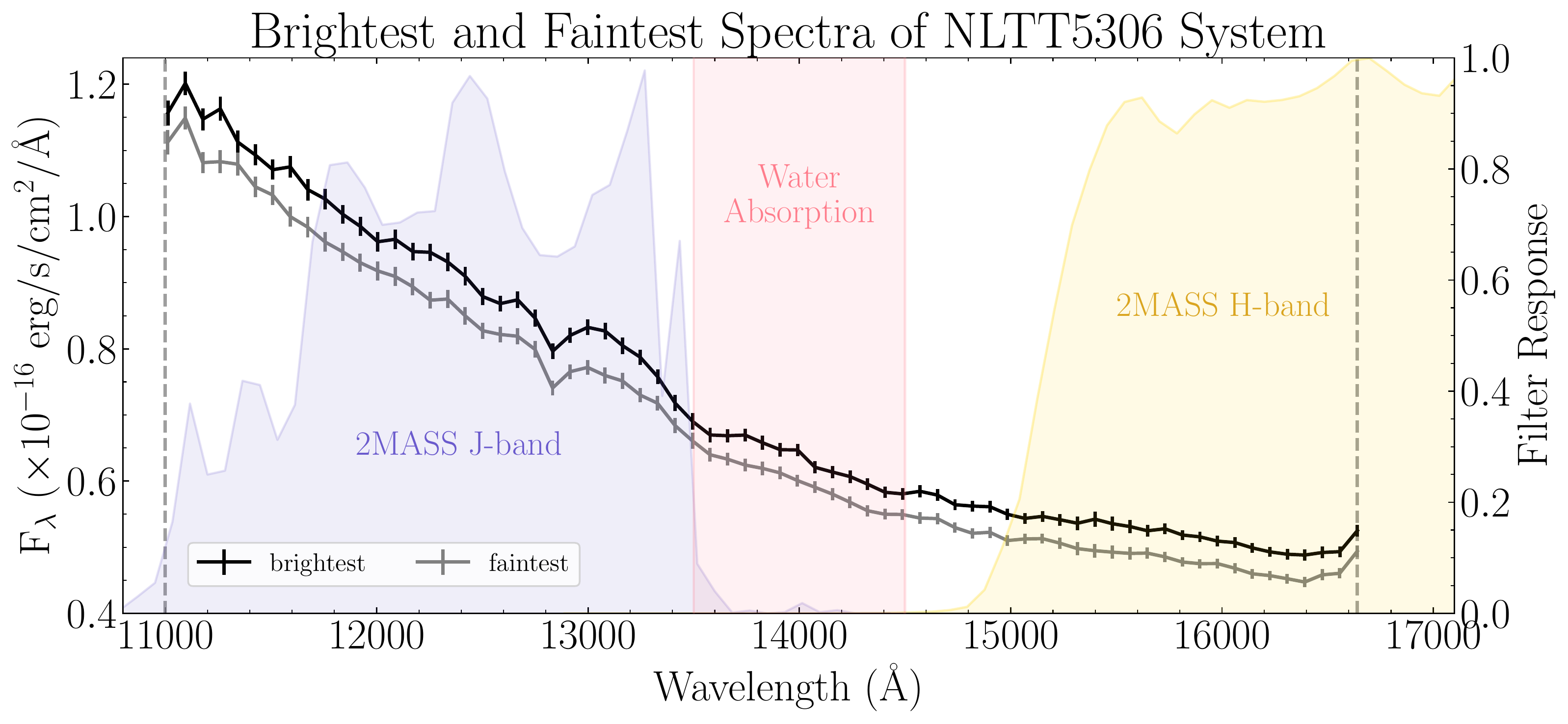}
\caption{Science spectra of NLTT5306, after data reduction steps, including ramp effect corrections, resampling, and wavelength cutoffs (vertical dashed lines; described in Section~\ref{sec:dataredux}). Filters profiles, shown in purple, pink, and yellow, were used to make synthetic light curves for better comparison with other observations. They were re-binned to the resolution of our data and truncated according to our wavelength cutoffs.}
\label{fig:finalspec}
\end{center}
\end{figure}

We prepared to run the \texttt{aXe} tasks, \texttt{axeprep} and \texttt{axecore}, by downloading the necessary configuration and reference files that are stored and versioned in the aXe (hstaxe) WFC3 Extraction Example Cookbook\footnote{\url{https://github.com/npirzkal/aXe_WFC3_Cookbook}}. We then ran \texttt{axeprep} on the G141 spectroscopic images to perform background subtraction, exposure time normalization, and gain correction. For optimal background subtraction, we performed bad pixel corrections on the master sky image, \texttt{WFC3.IR.G141.sky.V1.0.fits}. The process was identical to the method described above, except the locations for bad pixel corrections in the sky image were determined by the DQ extension in the first G141 observation of each orbit. Finally, we extracted the 1D spectra with \texttt{axecore} using a four pixel radius extraction window. Examples of an extracted 1D spectrum are shown in the right-hand panels of Figure~\ref{fig:dispimage}.

The data exhibited signs of the instrumental systematic known as ``ramp effect'', predominantly caused by electron charge-trapping and delayed release in the detector. We corrected for this effect by fitting a \texttt{RECTE} ramp-profile model \citep{Zhou17} to the uncorrected broadband light curve. The model created the ramp profile based on eight parameters. Values for the six fixed parameters controlling trapping processes, E$_{\rm{tot}}$, $\eta$, and $\tau$, for both slow and fast populations were slightly updated from \citet{Zhou17} and presented in Table~\ref{tab:recte}. The two free parameters, E$_{0,s}$ and E$_{0,f}$, representing the number of trapped charges at the beginning of each orbit, were found with a custom Markov Chain Monte Carlo (MCMC) performed by \texttt{emcee} \citep[][]{emcee}, detailed in Section~\ref{sec:mcmc}. The final ramp model was then divided out of each spectrum.

We resampled our spectra to the actual resolving power of our observations by modeling the spectral response function from each 2D~stamp output by \texttt{axeprep} and \texttt{axecore}. We plotted the counts of each pixel in the direction perpendicular to the dispersion axis and fit the shape of the count curves with a Moffat profile. We then converted the Full-Width at Half-Maximum values of the Moffat profiles from pixel-space to wavelength-space using the \texttt{CDELT} keyword in the direct image headers. The resulting resolution of the spectra was R$\sim$130 at 14,000~\AA{}.

The errors on the resampled spectra were calculated utilizing a method from \citet{Carnall17}. The new error, $\sigma_{\rm{new}}$, was estimated by 
\begin{equation}
    \sigma_{\rm{new}} = \frac{\sqrt{\sum\limits^{N}_{i=1} \sigma_i f_i}}{N d\lambda}
\end{equation}
where $\sigma_i$ is the error on the non-resampled spectrum, $f_i$ is the fraction of the old bin covered by the new bin, $N$ is the total number of old bins covered by the new bin, and $d\lambda$ is the resolution of the non-resampled data, which in our case, was $\sim47$~\AA.

We ensured that we were using quality data by making wavelength cutoffs wherever the resampled flux errors were larger than two times the average error value between 13,000 and 15,000 \AA. Using this condition, the resulting data on average spanned wavelengths between 11,010 to 16,670 \AA{} (Figure~\ref{fig:finalspec}).

\begin{figure}
\begin{center}
\includegraphics[width=0.46\textwidth]{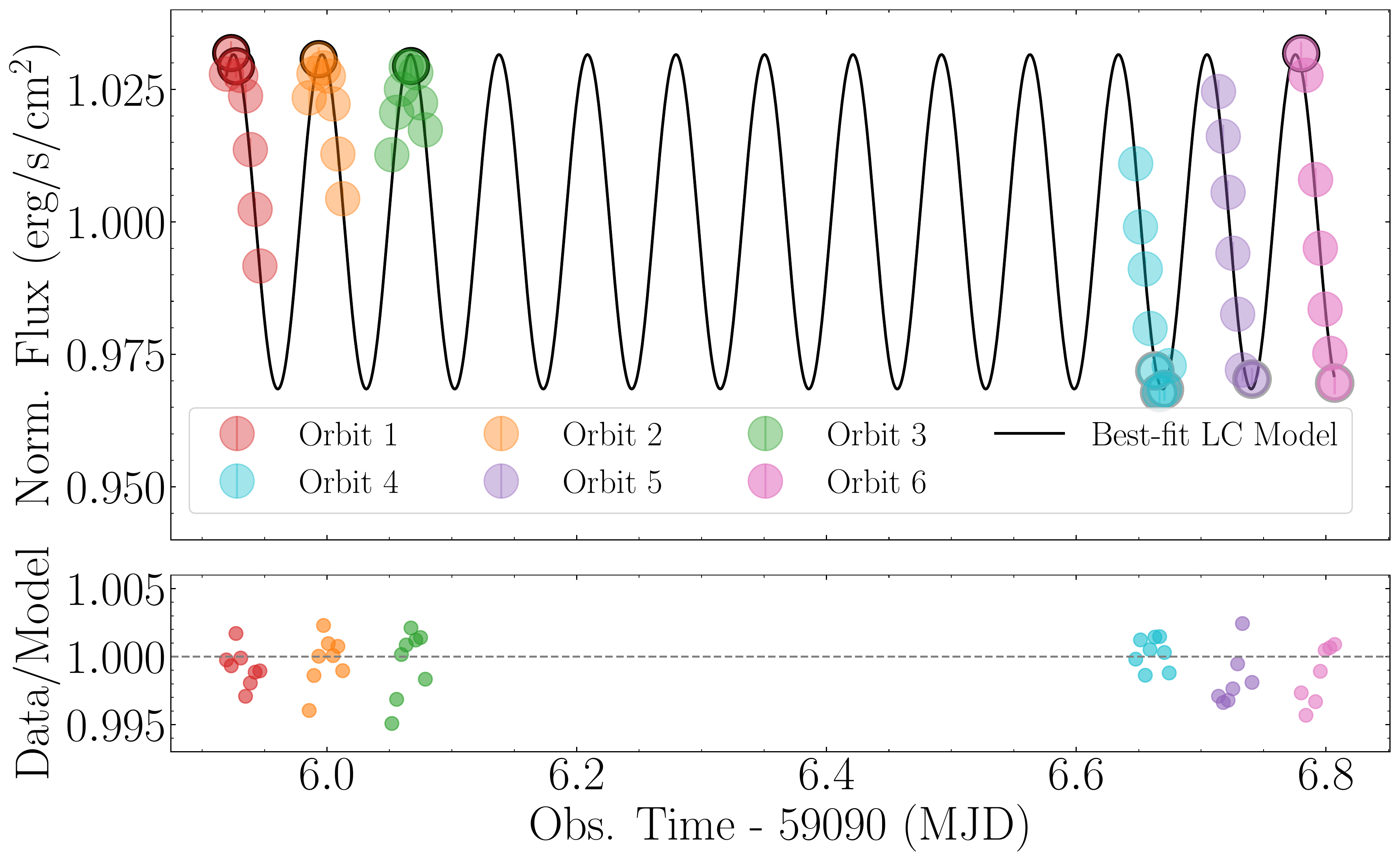}
\caption{Normalized broadband light curve of NLTT5306, using best-fit values for \texttt{RECTE} and phase curve parameters, described in Section~\ref{sec:mcmc} and presented in Table~\ref{tab:recte}. Using this approach, we found that a 1st-order phase curve adequately represented the data, bringing the residuals down below 0.5\%. Five brightest and faintest flux values are circled in black and gray, respectively.}
\label{fig:whitelc}
\end{center}
\end{figure}


\section{Light Curve Analysis} \label{sec:lcanalysis}

\subsection{Best-fit Light Curve and RECTE Parameters} \label{sec:mcmc}
For this data set, we simultaneously fit \texttt{RECTE} parameters (E$_{\rm{0,s}}$ and E$_{\rm{0,f}}$) and phase curve parameters. The number of phase curve parameters varied from three to five, depending on the order of the sine curve. The period ($P$) and amplitudes ($a_k$ and $b_k$) are free parameters, following methodology from \citet{Zhou22}. Basically, we treated the phase curves as a Fourier series, i.e., a combination of sinusoidal functions,
\begin{equation}
    \label{eq:combo_sines}
    F(t) = 1 + \sum_{k} a_{k} \sin (k \frac{2 \pi t}{P}) + 
    b_{k} \cos (k \frac{2 \pi t}{P}),
\end{equation}
where $F(t)$ is the phase curve as a function of observation time, $P$ is the period, $a_k$ and $b_k$ are the amplitudes for sine and cosine, and $k$ is the order of the sinusoidal function. So, for a 1st-order phase curve, the number of fitting parameters for the phase curve is three ($P$, $a_1$, and $b_1$), whereas this number increases to five for a 2nd-order phase curve ($P$, $a_1$, $b_1$, $a_2$, and $b_2$). Amplitudes were calculated following Equation (2) in \citet{Zhou22}, defined as: amp$_{k} = \sqrt{a_{k}^{2}+b_{k}^{2}}$.

\begin{deluxetable}{crcr}
\tablecaption{Updated RECTE Parameters \label{tab:recte}}
\tablehead{
\colhead{Parameter} & \colhead{Value} & \colhead{Parameter} & \colhead{Value} 
}
\startdata
\hline
\multicolumn{4}{c}{Fixed Model Values}\\
\hline
$E_{\rm{f,tot}}$ &  225.7 & $E_{\rm{s,tot}}$ & 2192 \\
$\eta_{\rm{f}}$ & 0.0116 & $\eta_{\rm{s}}$ & 0.02075 \\
$\tau_{\rm{f}}$ & 3344 & $\tau_{\rm{s}}$ & 16,300 \\
\hline
\multicolumn{4}{c}{Best-fit MCMC Values}\\
\hline
$E_{\rm{f,0}}$(1) & 4.95 $\pm$ 0.04 & $E_{\rm{s,0}}$(1) & 0.01 $\pm$ 0.00 \\
$E_{\rm{f,0}}$(2) & 50.24 $\pm$ 0.09 & $E_{\rm{s,0}}$(2) & 500.16 $\pm$ 0.04 \\
\enddata
\tablecomments{Model parameters for the \texttt{RECTE} ramp-profile model. In the top section, these values were treated as globally fixed for both the fast and slow populations. In the bottom section, these values represent the number of trapped charges at the beginning of each orbit and were allowed to vary in the MCMC run. We treated the two visits separately, where Visit~1 denoted as (1) and Visit~2 denoted as (2).}
\end{deluxetable}

\vspace{-0.3 cm}
When finding the \texttt{RECTE} parameters, we fit the two visits separately, adding four free parameters in our light curve model. The two visits are separated by over 8 \textit{HST} orbits, during which the charge trap status was not traceable and cannot be modeled by \texttt{RECTE}. Therefore, separately fitting the initial trapped charges are necessary and significantly reduces the ramp effect systematics.

To summarize the custom MCMC function, we first started by applying the \texttt{RECTE} parameters to the separate visits. We then created a broadband light curve from the \texttt{RECTE}-corrected spectra, normalized the light curve to the mean value between the 5 brightest and faintest points, and then created a phase curve model. Lastly, we calculated a likelihood value between the normalized light curve and model phase curve. The likelihood function was:
\begin{equation}
    \rm{ln}(\mathcal{L}) = 
    \sum -0.5\left( \frac{O - M}{\sigma_O} \right)^2 - \rm{ln}\left(\sqrt{2 \pi \sigma_O^2} \right)
\end{equation}
where O is the derived phase curve from spectral observations, M is the phase curve model, and $\sigma_{\rm{O}}$ is the error after integrating over wavelength. This process was repeated for an MCMC chain of $N_{\rm{Steps}}$=5,000 and $N_{\rm{Walkers}}$=2$\times N_{\rm{parameters}}$.

At the end, the best-fit values and uncertainties were derived from the chains after a burn-in of 250 Best-fit values for the \texttt{RECTE} parameters are presented in Table~\ref{tab:recte}. In Figure~\ref{fig:whitelc}, we present the final normalized broadband light curve, corrected for ramp effect, along with the best-fit phase curve model. Using this simultaneous MCMC approach, we were able to achieve residuals of less than 0.5\% between the data and model. From our trials, we found that a 1st-order Fourier series model adequately represented the data, with little to no improvement in the $\chi^2$ fits when using a 2nd-order model. Additionally, this proves the power of the physically-motivated ramp-effect model \texttt{RECTE}.

\begin{figure}
\begin{center}
\includegraphics[width=0.47\textwidth]{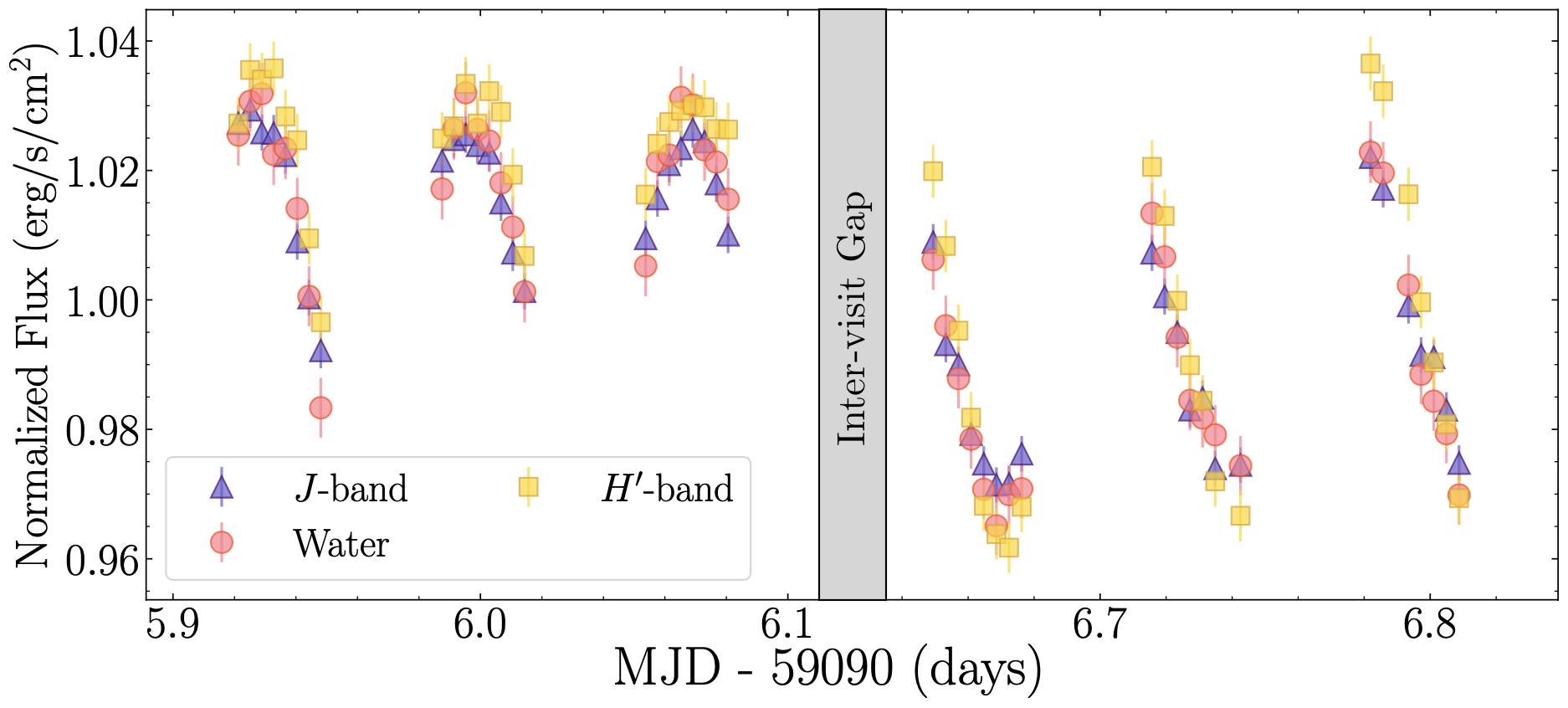}
\caption{Normalized synthetic light curves from the 2MASS $J$-, Water Absorption, and 2MASS $H'$-bands, integrated between 1.1 - 1.35, 1.35 - 1.45, and 1.5 - 1.67 microns, respectively. Our observations covered $\sim$6.1 periods of the NLTT5306.}
\label{fig:syntheticLCs}
\end{center}
\end{figure}

\begin{figure}
\begin{center}
\includegraphics[width=0.45\textwidth]{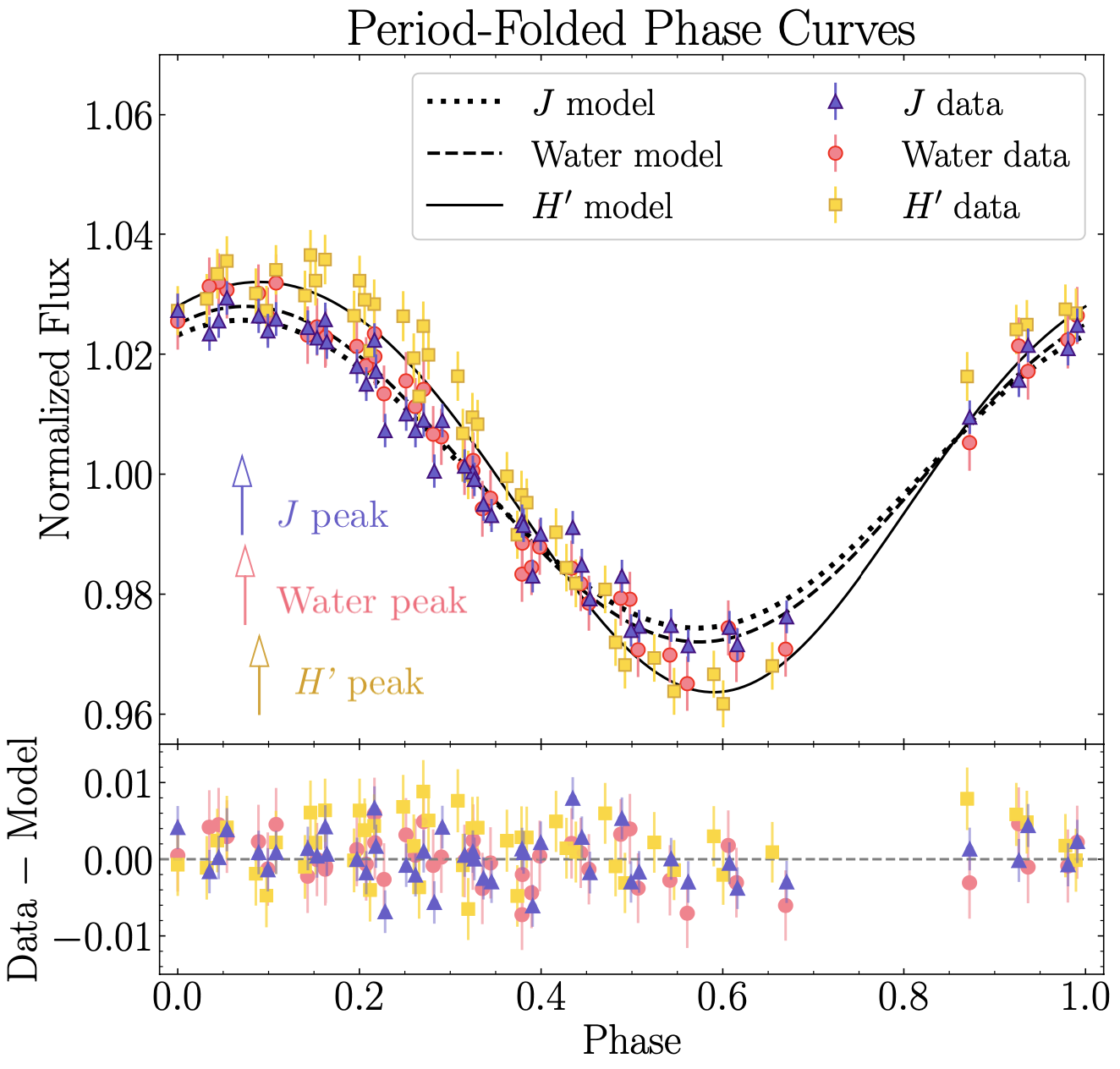}
\caption{Phase-folded light curves in the $J$ (indigo triangle), Water (pink circle), and $H'$ (golden square) bands along with best-fit phase curves as dashed, dotted, and solid lines, respectively. The peak of each filter is shown by arrows and labeled in the same color as the filter data. A phase of 0 here corresponds to first \textit{HST} observation. For the phase curve models, we checked for a dependence on \textit{HST} orbit and found none. In this figure, we observe a notable phase shift and higher amplitude in the $H'$ filter phase curve relative to the $J$- and Water bands.}
\label{fig:phasefold_lc}
\end{center}
\end{figure}

\subsection{Synthetic Light Curves}
To visualize and study intensity changes in a way that is easily comparable to similar analyses in other studies \citep[e.g.,][]{Buenzli12}, we derived synthetic light curves from the spectral observations. To create the synthetic $J$-band light curve, we convolved our spectra with the Two-Micron All Sky Survey (2MASS) $J$-band transmission filter from the 2MASS All-Sky Data Release \citep[][]{2MASS_Skrutskie06} across a wavelength range from 11,000~\AA{} to 13,500~\AA{}. We then integrated over these wavelengths and normalized by the mean flux. The same process was used to create the synthetic $H'$-band light curve, except we could not use the full wavelength range of the 2MASS $H$-band, since it extended beyond 16,700~\AA{}. Therefore, we created an $H'$-band light curve by integrating over 15,000 - 16,700 \AA{}. Finally, water absorption centered around $\sim$14,000~\AA{} is an important feature in atmospheric studies, so we also created a Water light curve by simply integrating from 13,500 to 14,500~\AA{}. The resulting synthetic light curves are presented in Figure~\ref{fig:syntheticLCs}.

\begin{deluxetable}{lrr}
\tablecaption{Phase Curve Parameters for synthetic $J$, Water, and $H'$-band Light Curves \label{tab:pcparams_JWH}}
\tablehead{
\colhead{Filter} & \colhead{Period} & \colhead{Amplitude}\\
\colhead{} & \colhead{[min]} & \colhead{[percent]}
}
\startdata
$J$-band & 101.88$\pm$0.06 & 5.131$\pm$0.43 \\
Water & 101.89$\pm$0.08 & 5.592$\pm$0.70 \\
$H'$-band & 101.92$\pm$0.02 & 6.990$\pm$0.31 \\
\enddata
\end{deluxetable}

\vspace{-0.5 cm}
In real atmospheres, different wavelengths tend to probe different pressures \citep[e.g.,][]{Buenzli12, Yang16} due to varying gas and dust opacities. Therefore, comparison of the light curves allows us to compare the brightness distribution in different atmospheric layers. For our three synthetic light curves, $J$-, Water, and $H'$-band, we created phase curve models to quantify any similarities or differences that could indicate pressure-dependent structure and dynamics. The models were created using a similar approach as described in Section~\ref{sec:mcmc}, except this time we did not need to correct for ramp effect, since ramp effect was already corrected for. Therefore, we only included phase curve parameters ($P$, $a_k$, and $b_k$) in this iteration of the MCMC function.

When fitting light curves for the $J$-, Water, and $H'$-band filters, we included higher-order models for comparison of $\chi^2$ values (i.e., $k$=1,\,2 and 4 in Equation~\ref{eq:combo_sines}), and found that the $J$- and Water band light curves were well fit $k$=1 models, whereas the $H'$-band light curve was better fit by a combination of $k$=1,2 models. One explanation for the $H'$-band light curve preferring a 2nd-order combination model, resulting in a broader peak and a narrower trough, could be a longitudinally confined area of flux emission \citep[e.g.,][]{Zhou22}, although it is not clear why the other filters would not show the same signature. The best-fit periods for the $J$-, Water, and $H'$-bands were 101.88$\pm$0.06 min, 101.89$\pm$0.08 min, and 101.92$\pm$0.02 min, respectively. The corresponding modulation amplitudes were 5.13$\pm$0.43 percent, 5.59$\pm$0.70 percent, and 6.99$\pm$0.31 percent. These values can also be found in Table~\ref{tab:pcparams_JWH}. The period-folded data and phase curve model for each filter are presented in Figure~\ref{fig:phasefold_lc}.

From Figure~\ref{fig:phasefold_lc}, we noticed a relative phase offset between the $H'$-band and the other light curves, highlighted by the horizontal offset between vertical arrows in Figure~\ref{fig:phasefold_lc}, which represent the peaks of each filter. We calculated the difference between the peaks for each light curve and found $H'-J$ = 6.80$\pm$0.03$^{\circ}$, $H'-$Water = 5.61$\pm$0.03$^{\circ}$, and Water$-J$ = 1.19$\pm$0.01$^{\circ}$. Additionally, we calculated the amplitude difference between the $H'$-band and other synthetic light curves to be over one percent, suggesting greater variability in flux at the pressures probed by the $H'$-band wavelengths.

\begin{figure*}
\begin{center}
\includegraphics[width=0.77\textwidth]{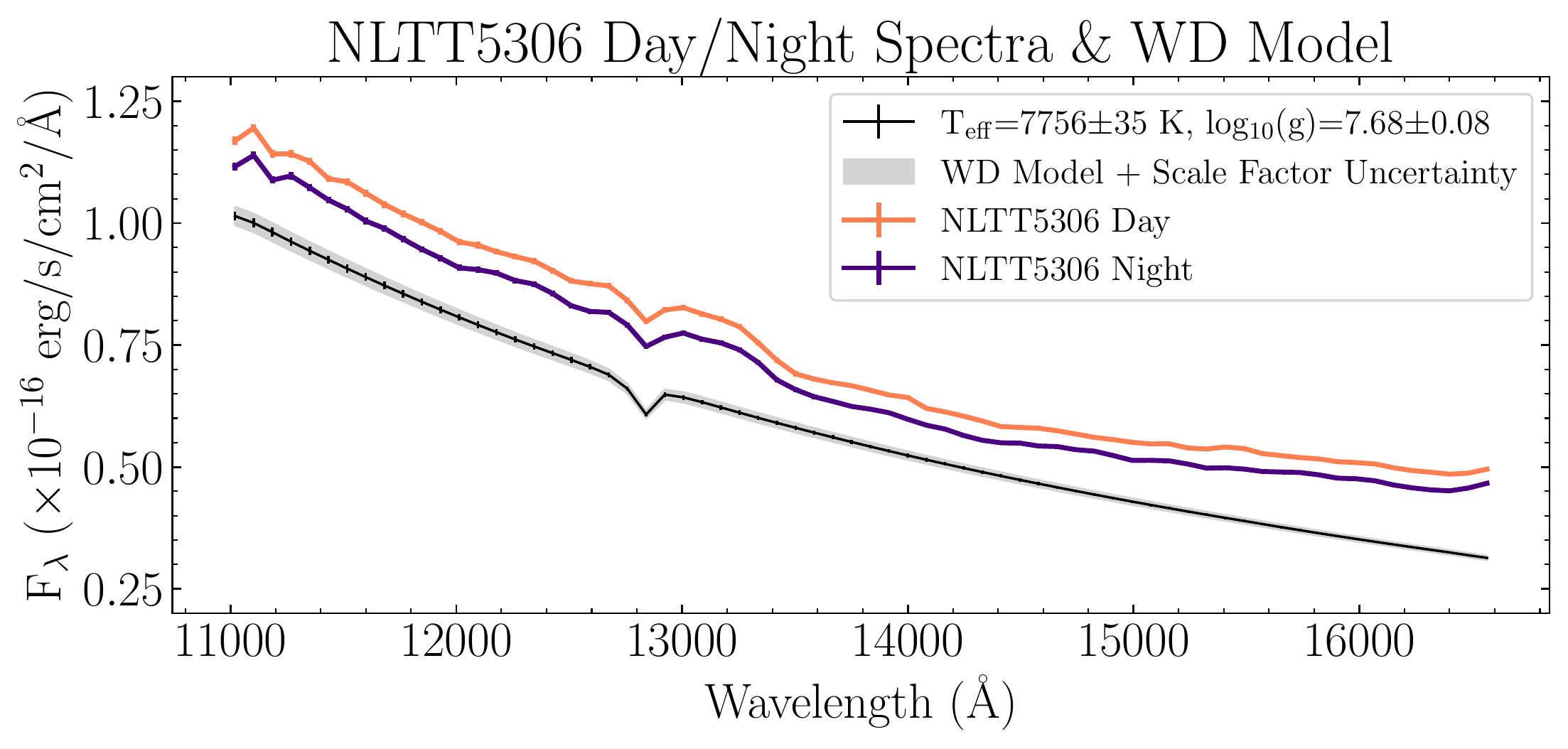}
\caption{The day- (orange) and night-side (purple) spectra of NLTT5306, corresponding positions A and B in Figure~\ref{fig:schematic}. The WD contribution is shown below in black. The uncertainties associated with the WD model, i.e. uncertainty in effective temperature and surface gravity, are shown as the black errorbars, whereas the total uncertainty, which includes uncertainty in scale factor (via uncertainty in distance to the system and white dwarf radius), is shown as the gray shaded region. From this figure, we can see that the contribution of the brown dwarf is approximately 20\% of the flux in the NIR.}
\label{fig:wdmodel}
\end{center}
\end{figure*}

Absolute phase offsets relative to eclipse would be useful for testing for evidence for any prograde (eastward) or retrograde (westward) hot spot shifts in the atmosphere of NLTT5306B. Doing so requires comparing the orbital phase of the brown dwarf to the phase of the maximum brightness. In the absence of an eclipse (which would provide orbital phase information), this could be achieved using precise radial velocity measurements. \citet{Longstaff19} calculated a radial velocity ephemeris using XSHOOTER data from 2014. However, the uncertainty on the period and orbital phase propagated to the time of our observations is too large ($\sim$10.1 hours or $\sim$6$\times$ the rotational period) to allow the precise determination of absolute phase offsets. Thus, current data is insufficient to test for a potential hotspot offset. In future observations, we encourage contemporaneous radial velocity and spectroscopic measurements.

\subsection{Investigating Phase Offsets and\\ Modulation Amplitudes}
Motivated by the $>$5 degree relative phase offsets and $>$1\% amplitude differences between the $H'$-band and the other synthetic light curves, we performed a more detailed analysis in an effort to pinpoint which wavelengths, and therefore which pressures, contributed most to these differences. Methods for this investigation are described in Appendix Section~\ref{sec:offs_amps}, with the results presented here.

In both the 1st and 2nd order models, the highest amplitudes reach $\sim$8\%, coming from two wavelength regions around 15,450 and 16,250~\AA{}, which would both fall into $H'$-band synthetic light curve. Following the $\sim$8\% peaks, there is a $\sim$7\% peak around 12,900~\AA{}, just inside the wavelength range of the $J$-band synthetic light curve. Phase offsets in the 1st order phase curve models (Figure~\ref{fig:offs_amps_k1}) mostly follow the same shape as the modulation amplitudes curve, except for the longer wavelength $H'$-band region, where the phase offsets appear to stay constant. In the 2nd order phase curve models, (Figure~\ref{fig:offs_amps_k2}), the phase offsets again follow the shape of the amplitude curve, except at the three wavelength regions where amplitudes exceed 7\%. Here, the offsets appear to be inversely correlated to the amplitudes peaks. The complex, highly wavelength-dependent behavior of the phase offsets and amplitudes are not seen in 3D models of irradiated brown dwarf atmospheres.

\section{Rotational Phase-Resolved Spectra of the Brown Dwarf NLTT5306B}
\label{sec:bdspec}

\subsection{Subtracting the White Dwarf Contribution} \label{sec:wdspec}
In order to obtain the phase-resolved spectra of the brown dwarf, it was first necessary to remove the contribution of the white dwarf from the observed WD+BD spectra. Since this system is unresolved and does not eclipse, it was not possible to get a spectrum of the white dwarf alone. Thus, we modeled the white dwarf spectrum using a model grid for pure-hydrogen atmosphere WDs from \citet{Koester10}, which used two free parameters, $T_{\rm{eff}}$ and log($g$). We adopted the published white dwarf values for these parameters (Table~\ref{tab:keyprops}) and bi-linearly interpolated along the \citet{Koester10} model grid. To account for uncertainty in the published values of $T_{\rm{eff, WD}}$ and log($g$)$_{\rm{WD}}$, we created 10,000 models for different values of temperature and surface gravity, sampled from Gaussian distributions with means of 7756 K and 7.68 cm~s$^{-2}$ and standard deviations of 35 K and 0.08 cm~s$^{-2}$, respectively. We then found the uncertainty of the WD model by measuring the FWHM of the fluxes from 10,000 models at each wavelength, seen as a solid black line with errorbars in Figure~\ref{fig:wdmodel}.

There was also uncertainty in the `scale factor' used to scale the model flux into what the expected observed flux would have been. To account for this noise source, we multiplied the flux at each wavelength by the scale factor, $\big(\frac{R_{\rm{WD}}}{D}\big)^2$, adopting published values and propagating uncertainties for $R_{\rm{WD}}$, white dwarf radius, and $D$, distance to the system. Therefore, the total error in the WD model comes from two sources, shown as the gray shaded region in Figure~\ref{fig:wdmodel}.

\begin{figure}
\begin{center}
\includegraphics[width=0.45\textwidth]{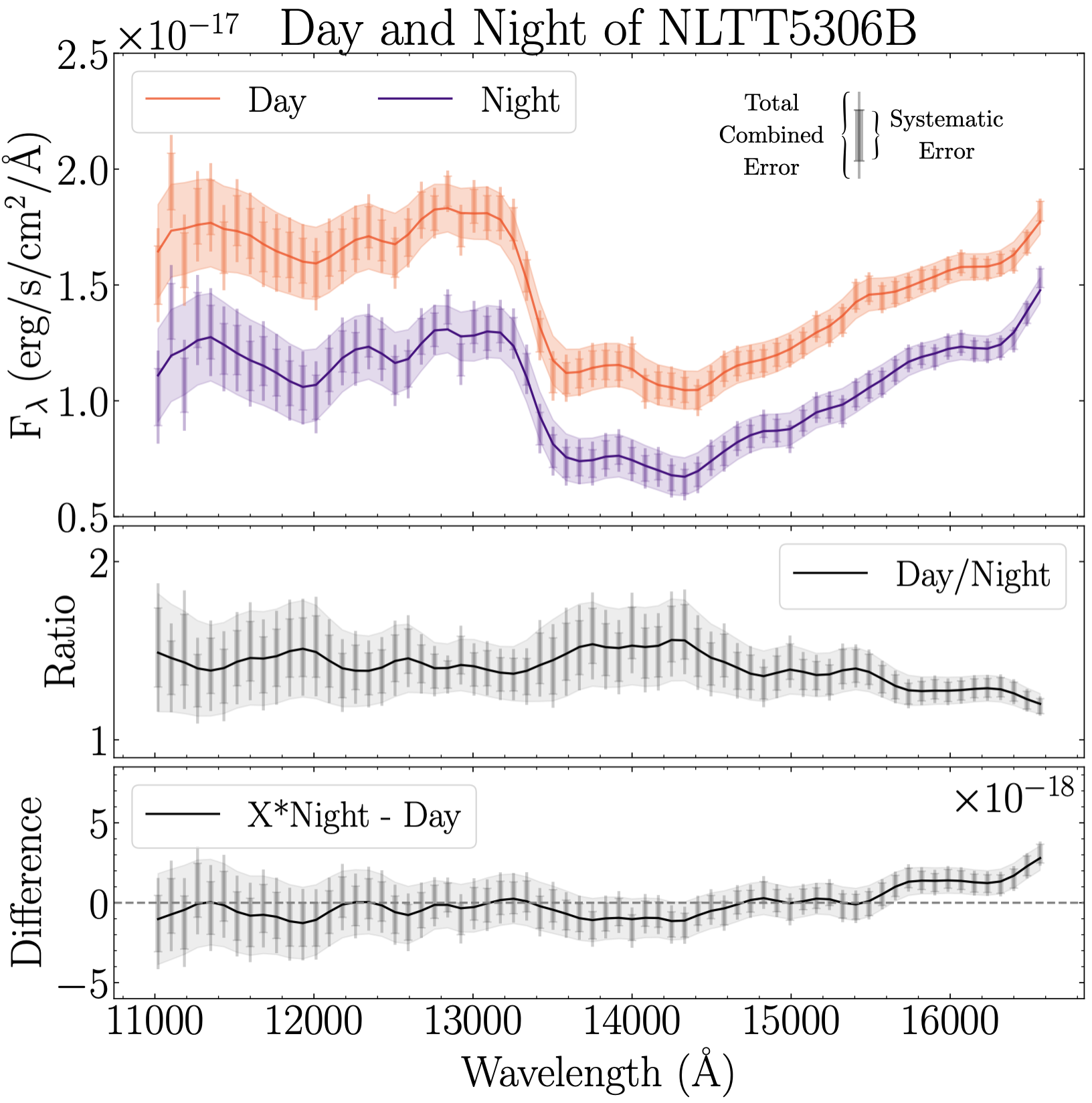}
\caption{\textbf{Top:} The Day/Night spectra of NLTT5306B. Systematic uncertainties and total combined errors are shown as thin and thick  erorrbars, respectively. Solid lines and shaded regions are observations and total uncertainties convolved with a Gaussian kernel with a width equal to the spectral resolution of our observations. \textbf{Middle:} The ratio between the day- and night-side spectra. Interestingly, while the overall ratio between the day and night spectra are roughly constant, there is a noticeable bump near the water absorption feature. \textbf{Bottom:} The difference between the day- and night-side spectra, where the night-side spectrum has been vertically scaled by X=1.37, such that the average of the difference is equal to zero. 
}
\label{fig:bdspecratios}
\end{center}
\end{figure}

\begin{figure}
\begin{center}
\includegraphics[width=0.47\textwidth]{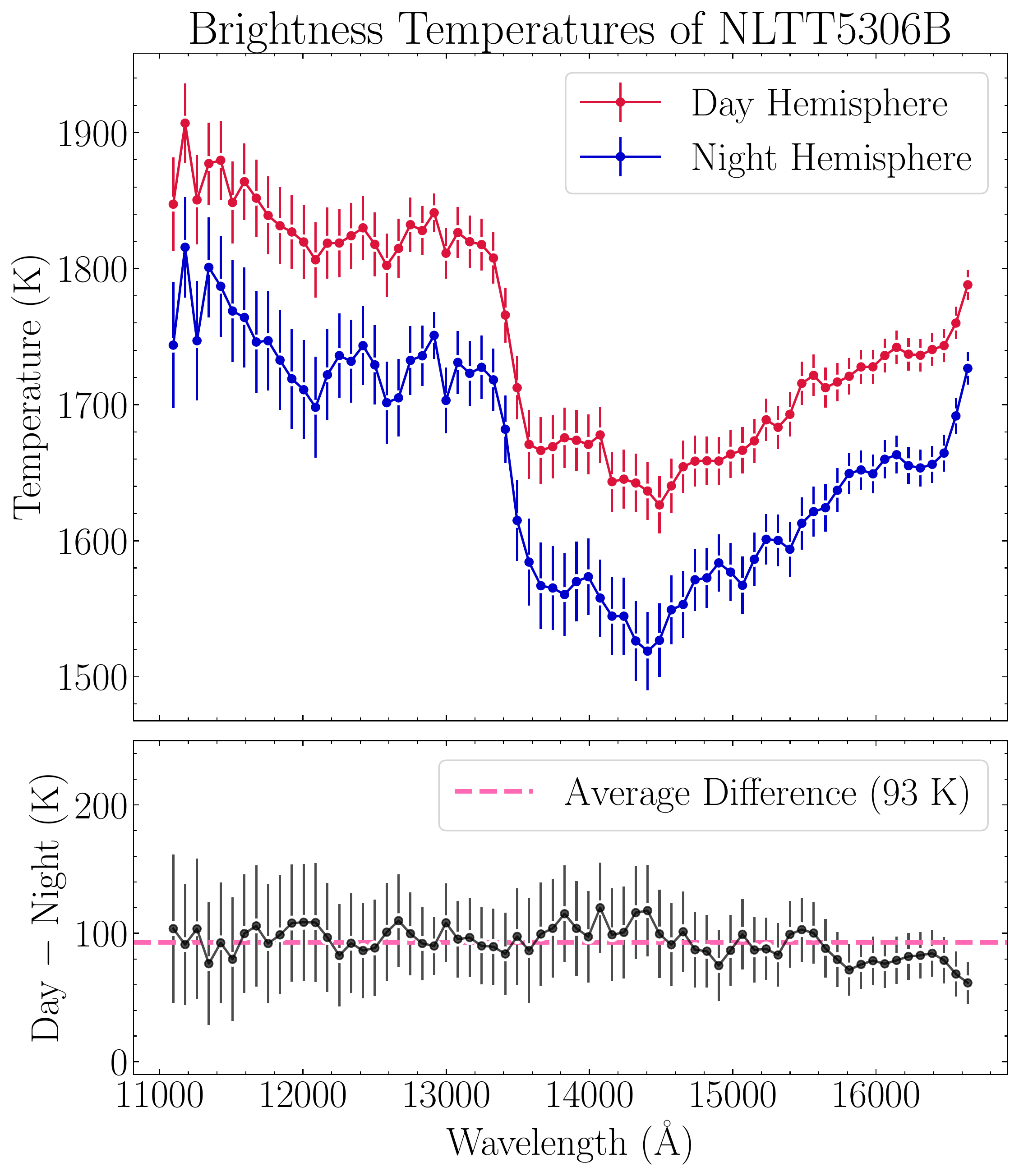}
\caption{\textbf{Top:} The day and night brightness temperatures of NLTT5306B, calculated by the solving the Planck equation for $T_{\rm{B}}$ at each wavelength. Errors here depict 1$\sigma$ uncertainties. \textbf{Bottom:} The difference in brightness temperatures between the day- and night-sides, along with the average difference over all wavelengths, plotted as a pink dashed line. Remarkably, the difference in brightness temperatures is nearly constant across the 1.1$-$1.67$\mu$m range.}
\label{fig:brighttemps}
\end{center}
\end{figure}

\subsection{Extracting Day/Night Spectra of NLTT5306B}
Thanks to our observations having near-complete phase coverage and very high data quality, we were able to confidently isolate and compare our analyses of the ``Day'' and ``Night'' sides of NLTT5306B. To further improve the signal-to-noise ratio, minimize any spectral variations, and secure the precision of the derived day/night spectra, we took the median of the five brightest and faintest spectra across all orbits and designated them as the day and night spectra. The corresponding brightest and faintest light curve points are circled in black and gray in Figure~\ref{fig:whitelc}. The NLTT5306 day- and night-side spectra produced with this method are shown in Figure~\ref{fig:wdmodel}.

\begin{figure*}
\begin{center}
\includegraphics[width=0.95\textwidth]{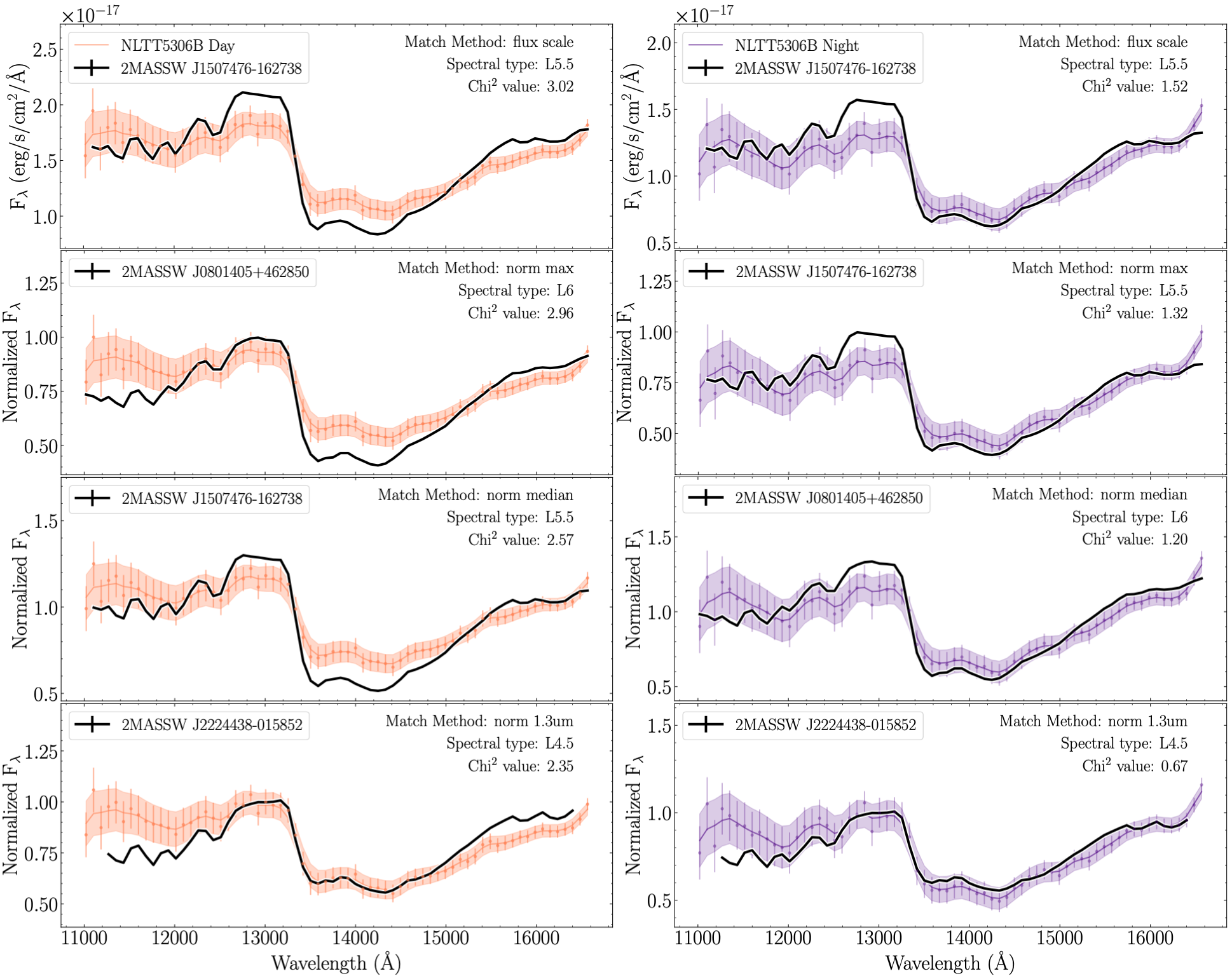}
\caption{Comparison of the day- and night-sides of NLTT5306B with their closest matching Cloud Atlas spectra. Plotting conventions for the observed brown dwarf spectra are the same as Figure~\ref{fig:bdspecratios}. Left column shows the day-side comparison (orange) whereas the right column shows the night-side (purple). Each row depicts a specific matching method. From top to bottom, they are: (1) Flux scaling to photometry, (2) normalizing to maximum flux value, (3) normalizing to median flux value, and (4) normalizing to flux at 13,000~\AA{}.}
\label{fig:cloudatlas}
\end{center}
\end{figure*}

Once the WD contribution was modeled and subtracted, we extracted the BD component for further atmospheric investigation. The isolated day- and night-side spectra are presented in the top panel of Figure~\ref{fig:bdspecratios}. The day-side is 40\% brighter than the night-side, on average, seen in the middle portion of Figure~\ref{fig:bdspecratios}. We attempted to rule out the possibility that the day-side is simply a direct scaling of the night-side by multiplying the night spectrum by a constant $X$=1.37 and then subtracting the day-side. From the non-constant shape of the difference curve, bottom portion of Figure~\ref{fig:bdspecratios}, we found that the day-night-side difference is not just in intensity, but also in their spectral features.

\subsection{Day/Night Brightness Temperatures}
\label{sec:brightnesstemps}
In the atmospheres of stellar companions, the combination of internal heat flux, profile of net absorbed stellar flux, and opacity structure control the thermal structure \citep[][]{MarleyRobinson15}. To study this structure in detail, we derived brightness temperatures at each wavelength, defined as the temperature at which a blackbody would emit as much specific intensity as the observed value. When coupled with a Pressure-Temperature (P-T) profile, brightness temperature is a useful proxy to map the relative pressure regions probed by each wavelength, which in turn reveals the vertical structure of the atmosphere at each observed phase. Using Planck equation and published values for brown dwarf radius and distance to the system (Table~\ref{tab:keyprops}), we calculated brightness temperatures versus wavelength for both the day-and night-sides of NLTT5306B, shown in Figure~\ref{fig:brighttemps}.

\begin{figure*}
\begin{center}
\includegraphics[width=0.85\textwidth]{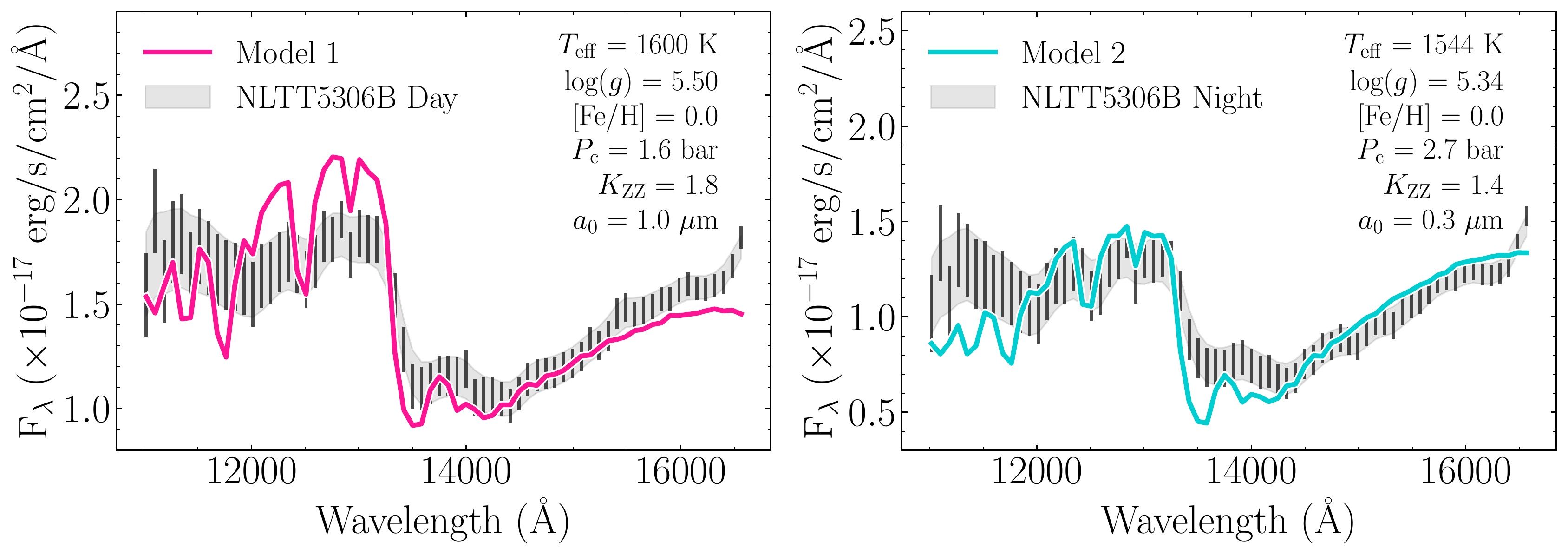}
\caption{Day (Left) and night (Right) spectra of NLTT5306B with best-fitting models from the cloudy, non-irradiated model grid that included cloud opacities from \citet{Brock2021}. Model parameters are labeled towards the right side of each figure. From these fits, we conclude that NLTT5306B most likely has an internal temperature around 1550 - 1600 K.} 
\label{fig:brockgrid}
\end{center}
\end{figure*}

In both hemispheres, the brightness temperatures showed strong wavelength dependence with temperatures that ranged between 1500 and 1900~K, with the lowest temperatures located around the water absorption band. These temperatures fall within the range where silicate clouds are expected to be an important opacity source \citep[][]{Ackerman01}. Since both the day and night hemispheres fall within this temperature range, it is likely that silicate clouds are present above the photosphere at all phases. The brightness temperatures also exhibited a nearly constant temperature difference of 93~K between day and night, further supporting the interpretation that both hemispheres possess similar gas-phase abundances and opacity sources. This small temperature difference was also seen in the P-T profiles presented in Figure~\ref{fig:pt_profile}.

\section{Comparison to Field Brown Dwarfs}
\label{sec:compare_fieldBDs}

\subsection{Cloud Atlas Spectral Library}
\label{sec:cloudatlas}
Next, to understand the impact of external irradiation on the atmosphere of NLTT5306B and brown dwarf atmospheres in general, we compare our derived irradiated brown dwarf spectra to a spectral library of non-irradiated brown dwarfs, also observed with the same telescope and instrument setup. Specifically, the Cloud Atlas Hubble Space Telescope Large Treasury program (PI: Apai) has, among other results, published a comprehensive, high-quality spectral library of 76 \textit{HST}/WFC3/NIR/G141 spectra. In order to investigate the differences between field brown dwarfs and irradiated brown dwarfs, we first sought to find a best-match spectral type from this library. Of the 76 Cloud Atlas spectra, 66 are spectra of L, T, and Y dwarfs (see Table~1 in \citealt{Manjavacas19}). We were able to use 53 of these spectra, since data for CD$-$352722b is missing from the downloadable tar.gz file and 13 others were excluded to ensure a high-quality data set with well-understood brown dwarf spectral types. The 13 spectra we excluded were: (1) eight objects were considered Planetary-Mass Companions (see Table~3 in \citet{Manjavacas19}), (2) three objects with two reported spectral types, and (3) one object with a spectral type of Y0pec.

Since there are uncertainties in many of the physical parameters we would use to find a best-match spectrum for the day- and night-sides of NLTT5306B, e.g. radius, effective temperature, and distance, we attempted to circumvent such uncertainties by using four different matching methods, three of which are close to being scale factor and temperature independent. The four methods, shown in Figure~\ref{fig:cloudatlas} for both the day and night-sides, included normalizing to the maximum flux value in wavelength range of our observations, normalizing to the median flux value, normalizing to the flux value at 13,000~\AA{}, and flux scaling to match photometry. The procedure for flux scaling is described in Appendix Section~\ref{sec:fluxscale_cloudtlas}.

Of the four matching methods used for the Cloud Atlas spectral library, normalizing to the flux value at 13,000~\AA{} provided the closest match of spectral features. This could be expected, since 13,000~\AA{} probes the highest pressure among the wavelengths considered here, and thus should be less affected by irradiation. For the day-side of NLTT5306B, the closest Cloud Atlas spectrum, with a $\chi^2$=2.35 was 2MASSW~J2224438-015852, an L4.5 dwarf \citep[][]{DupuyLiu12}. This brown dwarf was also the best-match for the night-side of NLTT5306B, with a $\chi^2$ value of 0.67. 2MASSW~J2224438-015852 is known to have a significant silicate cloud feature around 10 $\mu$m \citep[][]{Cushing06}, supporting our later interpretation of NLTT5306B as having significant cloud coverage.

\section{Non-Irradiated Brown Dwarf Models}
\label{sec:Compare_nonirr_models}

\subsection{Brock Model Grid}
\label{sec:BrockGrid}
In an effort to provide a baseline for the expected condensate clouds, we started our modeling procedure by comparing the observed day-side and night-side spectra of NLTT5306B to non-irradiated models that included cloud opacity from \citet{Brock2021}. These models had much the same setup as described in Section~\ref{sec:LothringerGrid}, but included the opacity from a variety of important condensate species, including silicates, corundum, iron, and many more \citep[][]{allard:2001}. The vertical extent of the clouds was determined by a vertical mixing parameter ($K_{\rm{zz}}$) and a $P_{\rm{c}}$ parameter, which describes the pressure at which the cloud particle number density begins to decrease exponentially. The mean particle size of the log-normal distribution of grain sizes was varied between 0.125 and 10~$\mu$m. 

To find a best fit model for day and night, we applied a scale factor, $\big(\frac{R_{\rm{BD}}}{D}\big)^2$, to all models and then calculated the $\chi^2$ value between each model and the day- and night-side spectra for the brown dwarf. We determined the best match for each side by finding the smallest $\chi^2$ value. Figure~\ref{fig:brockgrid} shows the best-fit cloudy, non irradiated models from the Brock model grid to the extracted day and night spectra of NLTT5306B, which are listed in full in Table~\ref{tab:models}. The models had $T_{\rm{eff}}$ between 1544 $-$ 1600 K and surface gravity, log($g$), values between 5.34 $-$ 5.50, for night and day, respectively. Along with the Cloud Atlas fits, we used the parameters of these models as a baseline for the brown dwarf parameters that the irradiated models would start from. As this system is not eclipsing, we chose to approximate the night-side of NLTT5306B as the closest estimate for a non-irradiated atmosphere. Thus, we concluded that the internal temperature of NLTT5306B was most likely between 1544 and 1600 K, meaning that without external irradiation, we could expect the dominant opacity sources to be H$_2$O, CO, and silicate grains, which are predicted for temperatures above 1500~K \citep[][]{Burrows01}.

\begin{figure}
\begin{center}
\includegraphics[width=0.47\textwidth]{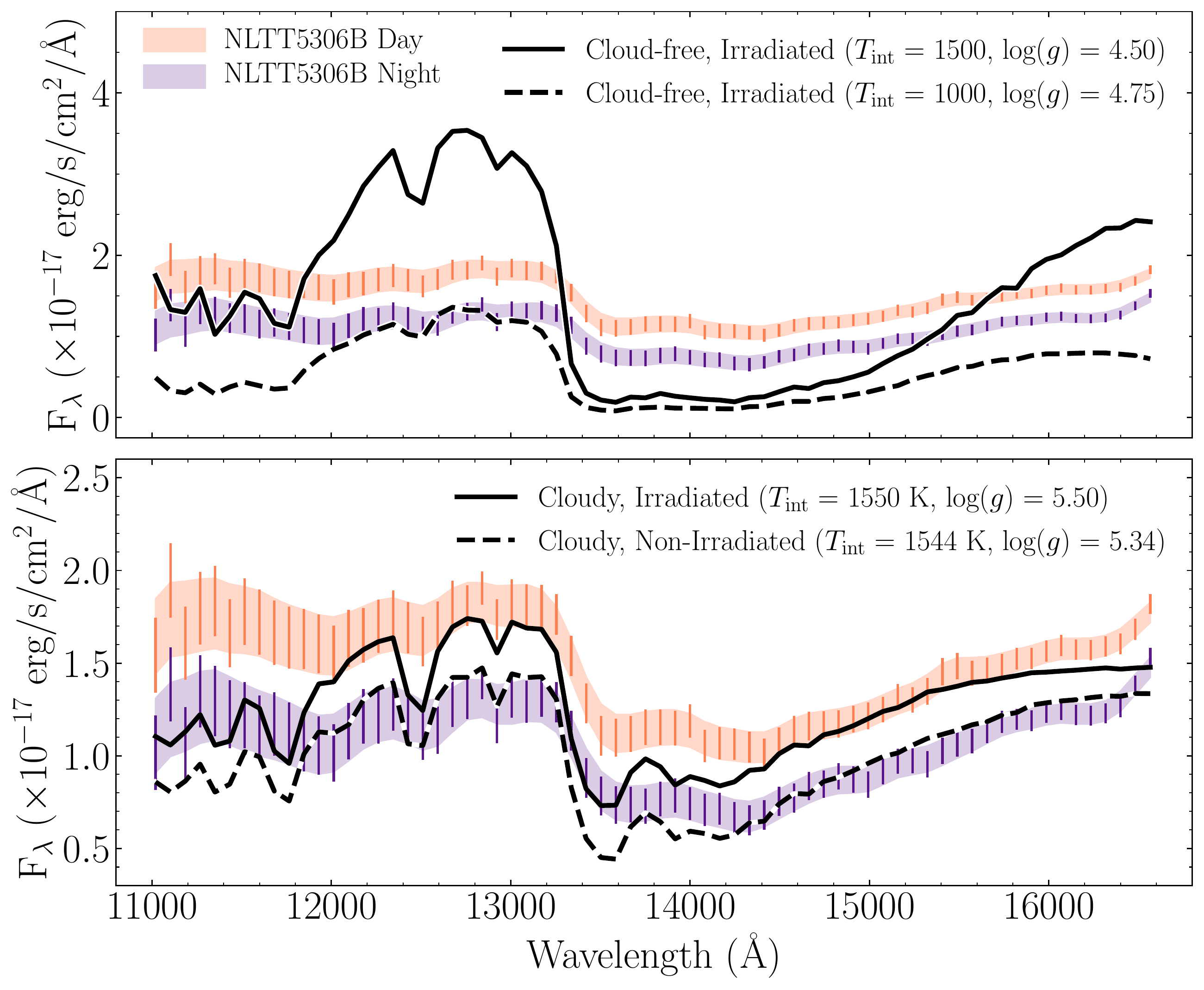}
\caption{\textbf{Top:} Comparison of day (orange) and night (purple) extracted spectra of NLTT5306B with best-fit cloud-free, irradiated models from the Lothringer Model Grid (described in Section~\ref{sec:LothringerGrid}). Neither model was a good match to the data. \textbf{Bottom:} Same as top panel, except with cloudy models where both irradiated and non-irradiated were considered.
The spectra of NLTT5306B are more muted than the models predict, however it is clear that they prefer cloudy over cloud-free models from this model grid.}
\label{fig:lothringermodels}
\end{center}
\end{figure}

\section{Irradiated Brown Dwarf Models }
\label{sec:Compare_irr_models}

\subsection{Lothringer Model Grid}
\label{sec:LothringerGrid}

We compared the observations to a grid of cloud-free PHOENIX irradiated brown dwarf atmosphere models, similar to those described in \citet{Lothringer_Casewell20}. The models are calculated from 10 to 10$^7$~\AA{} with 1-2~\AA{} sampling up to 5~$\mu$m, with coarser sampling at wavelengths longer than 5~$\mu$m. We varied internal temperature ($T_{\rm{int}}$=1000, 1500, 2000, 2500~K), surface gravity ($\log g$ = 4.5, 4.75, 5.0), metallicity ([Fe/H] = -0.5, 0.0, 0.5), and heat redistribution ($f$ = 0.125, 0.25, 0.5, where 0.5 implies day-side redistribution, 0.25 implies full heat redistribution, and 0.125 implies full redistribution with a non-zero albedo). The final effective temperature of the brown dwarf is a combination of the internal temperature ($T_{\rm{int}}$ --- temperature associated with internal heat flux) and temperature caused by external irradiation ($T_{\rm{irr}}$ --- see Equation~1 in \citet{Lothringer_Casewell20}), like so:
\begin{equation}
    T_{\rm{eff}} = \big( T_{\rm{int}}^4 + T_{\rm{irr}}^4 \big)^{1/4}
\end{equation} 

In models with no external irradiation, the effective temperature is the same as the internal temperature. The brown dwarf models use the BT2 H$_2$O line list \citep[][]{Barber2006}, as well as HITRAN 2008 for other major molecular absorbers \citep[][]{Rothman2009} and Kurucz data for atomic species \citep[][]{Kurucz1995}. We used the \citet{Koester10} white dwarf spectrum closest to NLTT5306A's fundamental parameters to irradiate the brown dwarf atmosphere.

Figure~\ref{fig:lothringermodels} shows the day and night spectra of NLTT5306B with four different models from the Lothringer model grids: two cloud-free and irradiated, one cloudy and irradiated, and one cloudy and non-irradiated. For the cloud-free and irradiated model grid, we used the same matching method as described in Section~\ref{sec:Compare_nonirr_models}, where we applied a scale factor and compared $\chi^2$ values. The resulting best matches, with all of their free parameters, are presented in Table~\ref{tab:models}. For the day-side, a $T_{\rm{eff}}$ = 1500~K, log($g$) = 4.50 model was the best match. However, the model's peak flux was almost two times higher than the observed peak flux and the model's water absorption feature was also much deeper. The best-fit for the night-side was a $T_{\rm{eff}}$ = 1000~K, log($g$) = 4.75 model. This model better matched the peak flux of the extracted night-side spectrum, although the water absorption feature was again too deep. 

The irradiated model grid was cloud-free, so poor model fits are not so surprising since we expect the atmosphere of NLTT5306B to be cloudy. As such, we also ran irradiated versions of the best-fitting Brock models, which included clouds, shown in the bottom panel of Figure~\ref{fig:lothringermodels}. With this method, we found that the best match for the day-side spectrum was a cloudy, irradiated model with $T_{\rm{eff}}$ = 1550~K and log($g$) = 5.50. For the night-side, a cloudy, non-irradiated model with $T_{\rm{eff}}$ = 1544~K and log($g$) = 5.34 provided the closest fit overall. This night-side match is the same model seen on the right side of Figure~\ref{fig:brockgrid}.

\begin{figure}
\begin{center}
\includegraphics[width=0.47\textwidth]{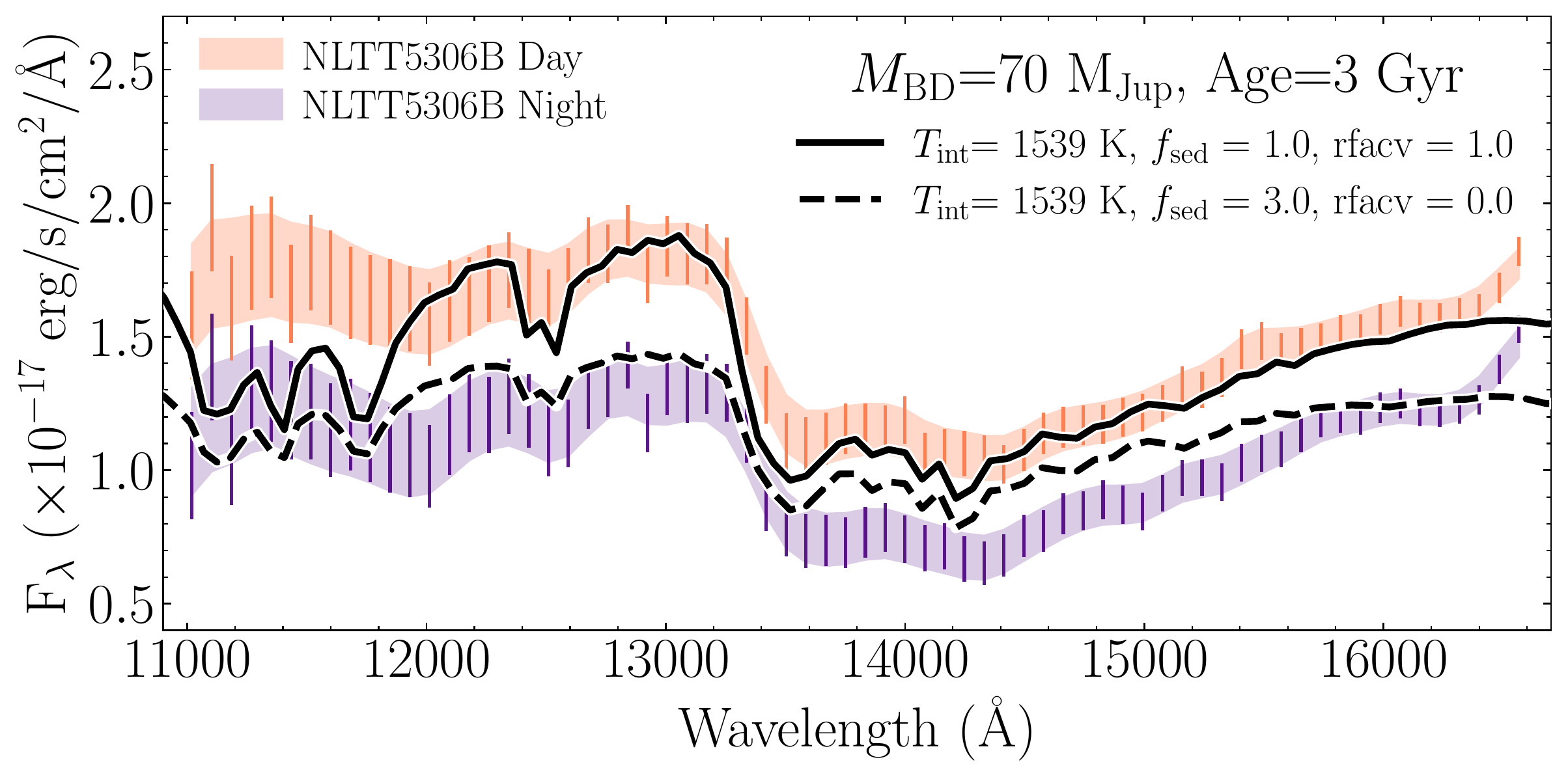}
\caption{Best-fit models from the Sonora model grid compared with extracted day and night spectra of NLTT5306B. The brown dwarf mass, age, and internal temperature are the same for both models, with the sedimentation and recirculation parameters ($f_{\rm{sed}}$ and rfacv) driving the difference between the two models.}
\label{fig:sonoramodels}
\end{center}
\end{figure}

\subsection{Sonora Model Grid} \label{sec:SonoraGrid}
We compared the observations to several forward models of both cloudless and cloudy atmospheres in 1D computed using the irradiated giant planet atmospheres code of \citet{Marley99} which generates an atmospheric radiative-convective equilibrium thermal profile based on given object properties,  based on the methods of \citet{McKay89} \citep[see also][]{Marley15}. We follow the same process employed by \cite{Mayorga19} and \cite{Lew22}, which we summarize here. 

We first computed cloud free models assuming an internal heat flux, parameterized as above by a $T_{\rm int}$, as determined by the \cite{Marley18} evolution grid and an age. We assumed chemical equilibrium and solar abundances. We did not include TiO and VO opacity in our calculation as these species are expected to be condensed into grains at these temperatures. 

\begin{figure}
\begin{center}
\includegraphics[width=0.45\textwidth]{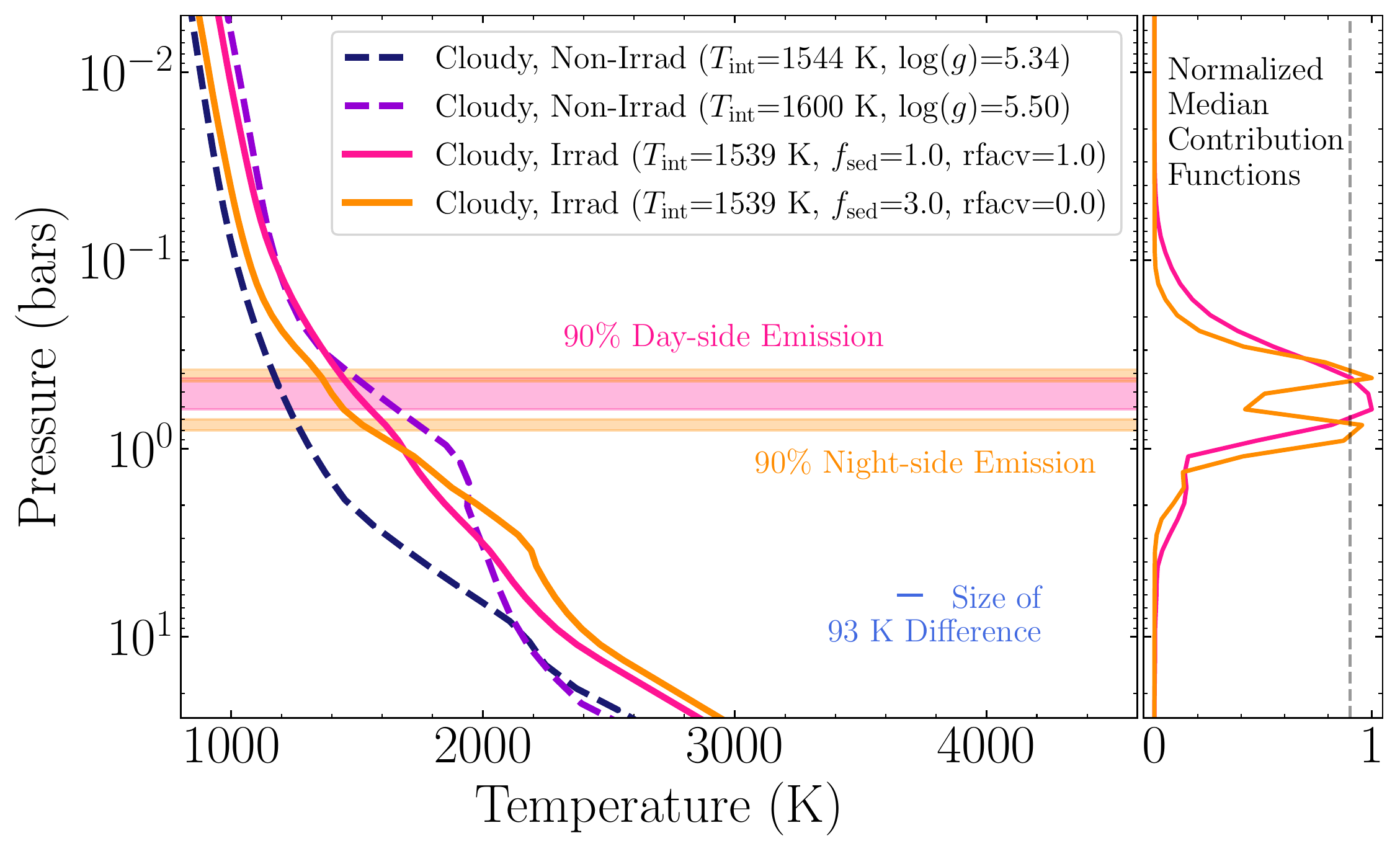}
\caption{\textbf{Left:} Pressure vs. temperature profiles for the models shown in Figures~\ref{fig:brockgrid} (dashed lines) and \ref{fig:sonoramodels} (solid lines). The size of the 93~K temperature difference seen in the observed day and night brightness temperatures (e.g. Figure~\ref{fig:brighttemps}) is shown as a blue bar in the bottom right. \textbf{Right:} Corresponding normalized contribution functions to the irradiated models. A gray dashed line at 0.9 depicts the pressure regions where $\geq$90\% of atmospheric emission originates for both models, also shown as shaded regions in the left panel.}
\label{fig:pt_profile}
\end{center}
\end{figure}

The high $T_{\rm{eff}}$ of the white dwarf primary (in excess of 7500\,K) results in a substantial fraction of its radiation being emitted blueward of the shortest atmosphere model spectral interval at $0.25~\mu$m. Because of the dearth of available UV opacities for gaseous absorbers the model cannot be straightforwardly extended to shorter wavelengths. To approximately account for the influence of the blue white dwarf flux on the model atmospheric structure, we applied a rudimentary correction by integrating the white dwarf flux short-ward of the bluest bin in the atmosphere model and added it uniformly to all of the spectral  bins short-wards of 0.4 $\mu$m (to avoid the potassium absorption lines). In essence this treats the bluest incident flux as having the same disposition as flux in the interval of 0.25 to $0.40~\mu$m.

To model the white dwarf primary we selected a \citet{Koester10} model at 7750~K and log($g$)=7.5 as the stellar host. We assumed a recirculation efficiency, rfacv, of 1 to represent a day-side average model rather than of 0.5 to represent a global average. We used PICASO \citep[][]{Batalha19} to generate the combined reflected and thermal emission spectra.

We also computed self-consistent cloudy 1D models for each companion allowing the following gas species to condense  as appropriate: NH$_3$, H$_2$O, KCl, ZnS, Na2S, MnS, Cr, MgSiO$_3$, Fe, Al$_2$O$_3$ \citep{Morley12, Marley13}. The particle size and vertical extent of the cloud layer was controlled by setting the cloud sedimentation efficiency parameter, $f_{\rm{sed}}$ \citep{Ackerman01}. 
A small $f_{\rm{sed}}$ yields tall lofted clouds of small particles while a large $f_{\rm{sed}}$ produces a vertically thin cloud with generally larger particles. Studies of brown dwarfs and Jupiter-like planets indicate that the appropriate $f_{\rm{sed}}$ to model clouds should be roughly 1--3 \citep{Stephens09, Ackerman01}. As with the cloudless models the profiles are iterated until a self-consistent radiative-convective equilibrium profile is obtained. We explored cloud free atmospheres and $f_{\rm{sed}}$ = \{3.0 and 1.0\}.

We first modeled the day-side using the given mass for the brown dwarf and age from \citet{Steele13} (52 M$_{\rm{Jup}}$ and 5~Gyr). In general, we found that these models were unable to provide a reasonable match to the observed spectra as they were far too cool. Our next step was to explore models with younger ages, which would increase their internal heat fluxes closer to those observed. Models with higher internal temperatures better matched the feature near 1.4~$\mu$m but, in general, lead to an over-prediction of the flux at all other wavelengths that could not be remedied by changing the cloud properties to sufficiently mute the features without masking them entirely. Varying the surface gravity of a model can also reduce the scale heights of features. Therefore, we explored models of 5~Gyr and 3~Gyr with masses ranging from 50-80~M$_{\rm{Jup}}$. We found that the best match of the observed spectrum was fit by a model with an internal temperature of 1539~K (3~Gyr, 70~M$_{\rm{Jup}}$) and $f_{\rm sed}$ = 1.0, shown in Figure~\ref{fig:sonoramodels} and listed in Table~\ref{tab:models}.

Using these best-fit parameters of 3~Gyr and 70~M$_{\rm{Jup}}$, we explored recirculation efficiencies, rfacv$<$1, to match the night-side spectrum with various cloud $f_{\rm sed}$ combinations. The night-side was best fit with an rfacv=0 and $f_{\rm sed}$ = 3.0, shown in Figure~\ref{fig:sonoramodels}. We found that it was difficult to match both the flux levels of both the short- and long-wavelength portions of the spectrum with the 1D model.

\section{General circulation modeling\\ of NLTT5306B} \label{sec:GCM}
To understand the general circulation and interpret the observed light curves of NLTT5306B, we adapted a general circulation model (GCM) appropriate for NLTT5306B. The atmospheric circulation and 3D thermal structure were simulated using the SPARC/MITgcm GCM which originated from simulating hot Jupiter atmospheres \citep{showman2009}. The model solves the global primitive equations of dynamical meteorology that are suitable for the outer most observable layers of giant planets and brown dwarfs. Radiative heating and cooling rates in each column of the GCM are calculated using the non-grey radiative transfer model of \cite{Marley99} that solves the two-stream radiative transfer equations and employs the correlated-k method. The correlated-k opacity tables have been updated based on the opacities used in \cite{Marley21_Sonora21}. We assume equilibrium chemistry in the atmosphere.

Here we introduce the specific setups of the GCM for NLTT5306B. The stellar spectra of NLTT5306 is obtained from the best fit used in Section~\ref{sec:wdspec} and the missing fluxes at lower and higher wavelengths are assumed to be blackbody radiation using the best-fit stellar temperature and radius. The temperature of the lowest model layer (at a pressure of about 130 bars) in the GCM is relaxed towards a prescribed value of 3700 K which is informed by the cloud-free Sonora grid model with a log($g\rm{)}=5$ and internal temperature of 1600 K. This boundary condition differs from previous setups of hot Jupiters (e.g., \citealp{showman2009}) in which a uniform net heat flux was prescribed. Our bottom boundary condition mimics a rapid entropy mixing by vigorous interior convection, and we argue that this is more appropriate here because convection mixes the entropy (towards the uniform interior entropy) rather than heat flux and that our GCM deep layers have reached the convective zone. To deal with the convective instability, we apply a convective adjustment scheme to remove convective instability instantaneously  (the implementation is described in \citealp{Tan2022}). The primary WD is much smaller than and very close to the BD companion, and we have applied a geometry correction of the WD irradiation  on the BD atmosphere. A linear drag is applied at pressures larger than 50 bars to crudely  represent interactions with the interior similar to that implemented in \cite{Tan2022}. The model domain extends from 150 bars to $3\times10^{-4}$ bar with 53 vertical layers and has a cubed-sphere C64 horizontal resolution (equivalent to $256\times128$ in longitude-latitude grid).

\begin{figure}
\begin{center}
\includegraphics[width=0.47\textwidth]{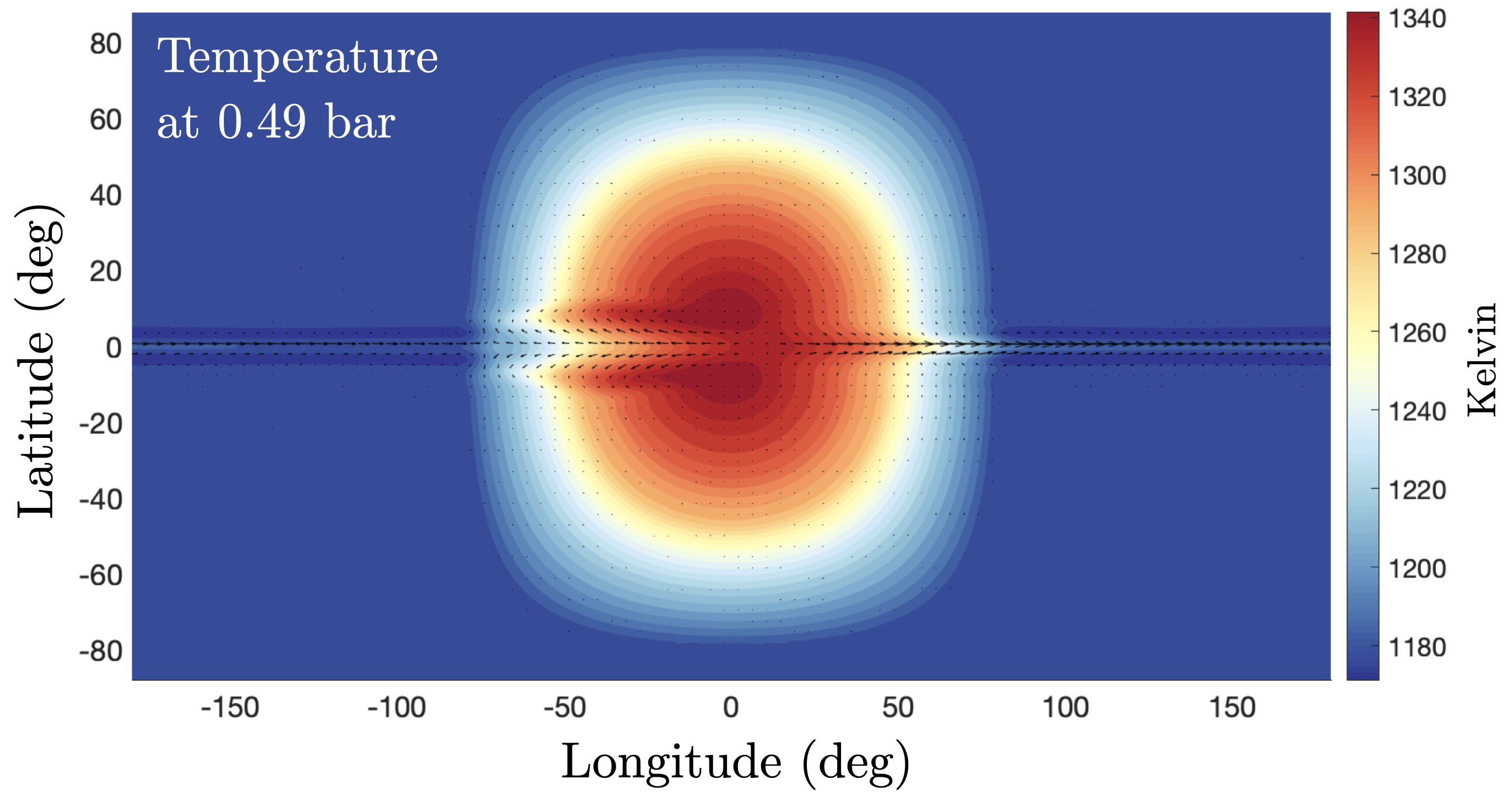}
\caption{Temperature distribution at $P$=0.49 bar in the GCM simulation of NLTT5306B. This pressure level was chosen since it is roughly in the middle of where most of the emitting flux originates from based on Figure~\ref{fig:pt_profile}. We see an eastward jet at the equator with thin westward-shifted lobes at equator-adjacent latitudes.}
\label{fig:cloudfree_gcm}
\end{center}
\end{figure}

The global temperature map at 0.49 bar from the GCM results of NLTT5306B is shown in Figure \ref{fig:cloudfree_gcm} along with the horizontal wind vectors as black arrows.  The most prominent feature is that the overall temperature pattern does not significantly deviate  from that determined purely by the radiative transfer, meaning that circulation is inefficient to drive strong winds in the global scale to shift the temperature pattern. The day-to-night temperature variation appears to be much smaller than the global mean temperature; this is not due to dynamical heat transport but rather is because the irradiation is much  weaker than the internal energy of NLTT5306B. Despite the global inactive circulation, a somewhat strong equatorial jet stream appears but is confined within a few degrees around the equator. At two sides of the equatorial jet, two meridionaly narrow hot lobes develop and are shifted westward; these are the so-called stationary Rossby lobes which are  rotational response to the day-night thermal contrast. The horizontal temperature structures at other pressures are qualitatively similar to that shown in Figure \ref{fig:cloudfree_gcm}. Such a circulation pattern is related to the rapid rotation of NLTT5306B as explained in \cite{TanShowman20_rotationWDBDs} and also shown in GCM results for other BDs around WDs \citep{Lee20,Sainsbury-Martinez21,lee2022}.

Because of the globally inactive circulation, the thermal light curves from the GCM results show negligible phase offsets. The wavelength-dependent thermal phase curves were generated by feeding the time-average global temperature structure to PICASO \citep{Batalha19,Robbins2022}, which shared the same opacity source as our GCM and so ensured consistency between post processing and the GCM. Phase curves were created for a sample of wavelengths representative of the \textit{HST} spectra, and phase curve models were fit using the same methods described in Section~\ref{sec:offs_amps}. The amplitudes and relative phase offsets were calculated for every model and investigated for wavelength-dependent behavior. Unlike what it seen in the observations (e.g. Figures~\ref{fig:phasefold_lc}, \ref{fig:offs_amps_k1}, and \ref{fig:offs_amps_k2}), there are no relative phase offsets among the thermal phase curves from the adapted GCM. Model amplitudes were strongly dependent on wavelength (ranging from 2 to 40\%), in contrast to the smaller amplitude changes seen in the observed phase curves.

Our cloud-free GCM practice suggests that the  cloud-free circulation scenario cannot explain the observed light curves. This is not surprising because clouds are already shown to be essential to interpret the observed day-side and night-side spectra (see section \ref{sec:Compare_nonirr_models} and \ref{sec:Compare_irr_models}). The importance of clouds and internal heat flux points to two possible improvements. The first is to consider the mechanism generating light curve variability of many isolated BDs by radiatively cloud driven circulation \citep{Tan2021}. The second is related to waves triggered by interactions between convection and the overlying stratified layers which could result in vertically sheared zonal flows in both isolated BDs \citep{showman2019,Tan2022} and young hot Jupiters \citep{lian2022}. Of course, mechanisms of cloud radiative effects in hot-Jupiter-like irradiation (e.g., \citealp{roman2019,parmentier2021,Komacek2022}) would also be important when interpreting cloud models for these type of objects as irradiation is still nontrivial.

\section{Discussion} \label{sec:disc}

\subsection{System Parameters}

Here we compare the newly calculated light curve period of NLTT5306 with values from previous studies. The first published period of $P$=101.88$\pm$0.02 min was determined via radial velocity (RV) observations back in 2013 by \citep[][]{Steele13}, 8 years before these \textit{HST} observations. \citet{Longstaff19} then constrained the period further to $P$=101.88 min $\pm$ 0.000002 ms in 2019. While 8 years is short compared the timescale of orbital decay \citep[approximately 900 Myr;][]{Steele13}, we might still expect to find a shorter orbital period, as this system is experiencing orbital decay on its way to becoming a cataclysmic variable.

Our best-fit broadband light curve model of NLTT5306 resulted in a period of 102.05$\pm$0.07 min, longer than the published values from \citet{Steele13} and \citet{Longstaff19}. Including uncertainties, the value we found is just over 3$\sigma$ away from those published values. This could be due to the fact that radial velocities trace orbital motion, whereas phase curve observations trace atmospheric rotation. \citet{Zhou22} also found a slightly longer light curve period than the radial velocity period of WD~0137. Therefore, if NLTT5306B’s atmosphere contains dynamics that exhibit local retrograde motion, like a westward traveling feature, the period measured by the phase curve can be prolonged even further. Thus, we cannot yet definitely say if the orbital period has evolved.

One system parameter that may impact our findings is the uncertain orbital inclination of NLTT5306. While precise orbital inclination was unknown, we were able narrow down the possible range of values based on observational constraints. For example, due to the lack of any observed transit in the phase curve, \citet{Buzard22} previously estimated an upper limit of $i \leq 78.7^\circ\pm0.4^\circ$ with the following equation:
\begin{equation}
    i_{\rm{partial}} = 90^{\circ} - \rm{sin}^{-1}\Big(\frac{R_{\rm{BD}} + R_{\rm{WD}}}{a}\Big),
\end{equation} 
and using the brown dwarf radius (informed by evolutionary models) of $R_{\rm{BD}}=0.095\pm0.004$~R$_{\odot}$ (Table~\ref{tab:keyprops}). Additionally, we used the observed radial velocity semi-amplitude and its connection to the two masses in the system, and the inclination as follows:
\begin{equation}
    \frac{M_2^3 \rm{sin}^3(i)}{( M_1 + M_2) ^2} = \frac{PK^3}{2\pi G},
\end{equation} 
where $M_{\rm{1}}$ is white dwarf mass, $M_{\rm{2}}$ is brown dwarf mass, $i$ is inclination, $P$ is orbital period, and $K$ is semi-amplitude of radial velocity. Although the mass of the brown dwarf is not directly measured, we may use $M_{\rm{2}}$ = 75~M$_{\rm{Jup}}$ as a conservative upper limit. With this, solving for $i$, and adopting values from Table~\ref{tab:keyprops} for $M_{\rm{1}}$ and $P$,  and $K_{\rm{H_{emis}}} = -49.1\pm1.1$~km/s from \citet{Longstaff19}, we found a lower limit of $i \geq 58.0\pm1.1$ degrees. Thus, for NLTT5306, we estimate the possible range of inclinations to be between between 58$^{\circ}$ and 79$^{\circ}$. In the following section (and in \S~\ref{sec:other_wdbds}), we discuss how this range of possible inclinations could impact our results.

\subsection{Longitudinal Temperature Distribution} \label{sec:longtempdist}
\begin{figure}
\begin{center}
\includegraphics[width=0.47\textwidth]{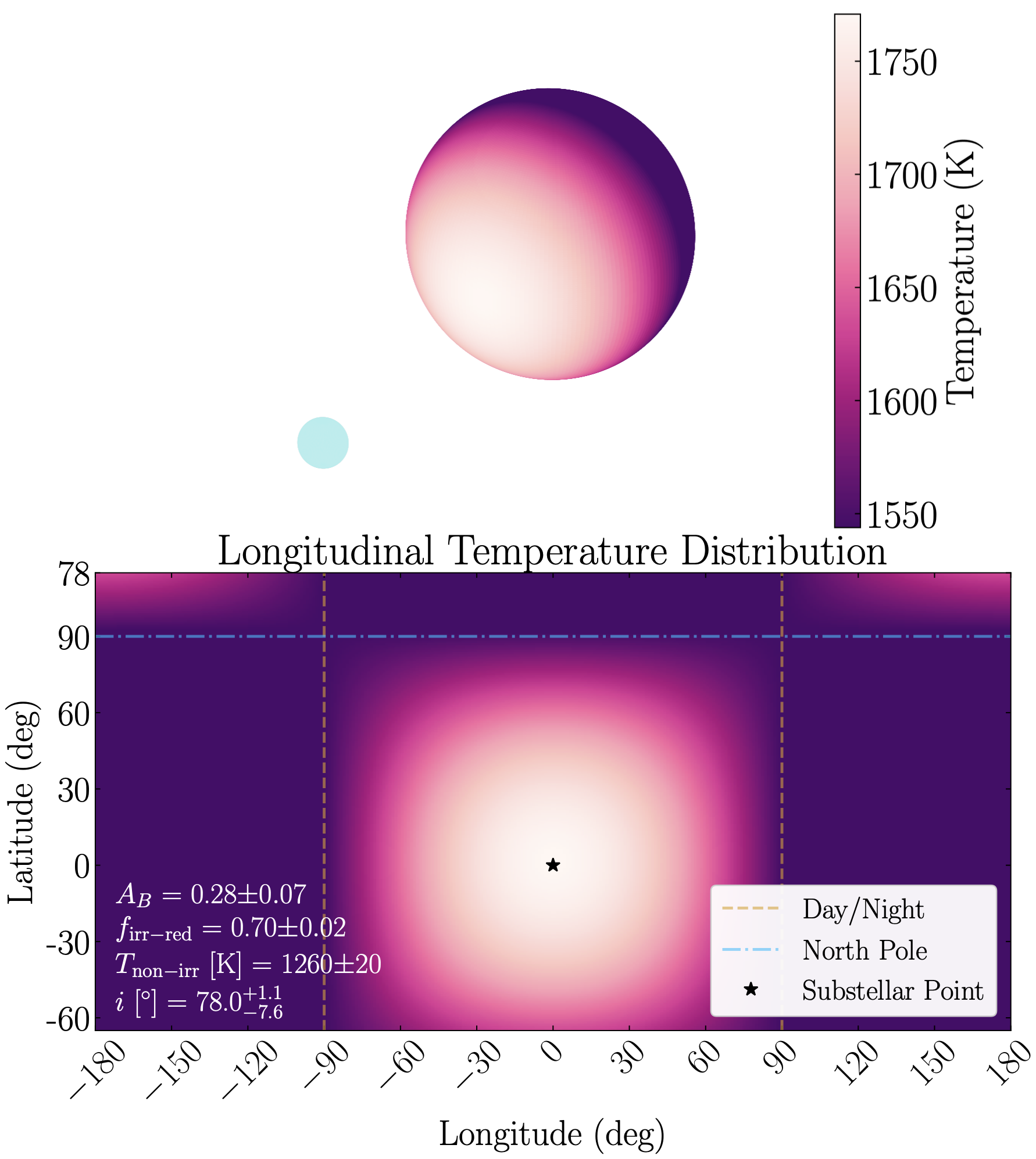}
\caption{\textbf{Top: } A 3D depiction of the temperature distribution on NLTT5306B. External irradiation comes from the WD primary, represented as a pale blue dot in the lower left. \textbf{Bottom: } A 2D projection of the 3D toy model above with labeled values for Bond albedo, irradiation redistribution fraction, non-irradiated BD temperature, and inclination. The day and night hemisphere-integrated temperatures have a difference of 93 K, matching derived temperature difference in Figure~\ref{fig:brighttemps}.}
\label{fig:radequilmap}
\end{center}
\end{figure}

With the known observed parameters of the system, (i.e., orbital separation and temperatures and radii of both dwarfs), we constructed a simple radiative equilibrium-based temperature distribution map of NLTT5306B, shown in Figure~\ref{fig:radequilmap}. The calculation of this map consisted of a grid search among four parameters: (1) Bond albedo ($A_B$), which set the fraction of incident irradiation that is reflected away from the day-side, (2) irradiation redistribution fraction ($f_{\rm{irr-red}}$), which controlled how much of the non-reflected irradiation was redistributed to the night-side, (3) non-irradiated brown dwarf temperature ($T_{\rm{non-irr}}$), which set the base temperature of the brown dwarf without irradiation, and (4) Inclination ($i$), which effectively controlled how much of one hemisphere was visible when the other hemisphere was dominating the emission. The grid search process is further described in Appendix Section~\ref{sec:tempmap_gridsearch}.

Figure~\ref{fig:radequilmap} shows the 2D temperature distributions from the best-fit combination of parameters: $A_B$ = 0.28$\pm$0.07, $f_{\rm{irr-red}}$ = 0.70$\pm$0.02, $T_{\rm{non-irr}}$ = 1260$\pm$20 K, $i$ = 78.0$^{+1.1}_{-7.6}$ degrees. The parameters were fit to match a night-side temperature of 1500~K, informed by the night-side brightness temperature at 1.4~$\mu$m, and a day-side temperature of 1593~K, motivated by the 93~K difference seen in derived brightness temperatures between day and night (Figure~\ref{fig:brighttemps}). We note that the relatively high Bond albedo is physically consistent, given that most the incident radiation from the white dwarf is in the UV/Optical blue and the predicted global presence of silicate clouds in this atmosphere \citep[][]{Marley99}. Although this model is rather simple, our ability to obtain a good match to observations with physically reasonable parameters suggests that we are able to capture the fundamental processes at play. Such models are powerful in providing physical intuition and insights into the key processes within these atmospheres.

\begin{deluxetable*}{lrcc|lrcc}
\tablecaption{Best Fitting Models \label{tab:models}}
\tablehead{
\colhead{Parameter} & \colhead{Value}  & \colhead{Cloudy?} & \colhead{Irradiated?} & \colhead{Parameter} & \colhead{Value}  & \colhead{Cloudy?} & \colhead{Irradiated?}
}
\startdata
\hline
\multicolumn{4}{c|}{Sonora Model Grid} & \multicolumn{4}{c}{Lothringer et al. Models}\\
\cline{1-4}\cline{5-8}
\multicolumn{2}{l}{\textbf{Day-side Match}} & Yes & Yes & \multicolumn{2}{l}{\textbf{Model 1}} & No & Yes \\
$T_{\rm{eff}}$ $^*$ [K] & 1539 &  &  & $T_{\rm{eff}}$ [K] & 1500 &  & \\
$f_{\rm{sed}}$ & 1.0 &  &  & log($g$) [cm s$^{-2}$] & 4.50 &  & \\
rfacv & 1.0 &  &  & $[$Fe/H$]$ & 0.5 &  & \\
Age [Gyr] & 3 &  &  & $f$ & 0.1 &  & \\
$M_{\rm{BD}}$ [M$_{\rm{Jup}}$] & 70 &  &  &  &  &  & \\
\multicolumn{2}{l}{\textbf{Night-side Match}} & Yes & Yes & \multicolumn{2}{l}{\textbf{Model 2}} & No & Yes \\
$T_{\rm{eff}}$ [K] & 1539 &  &  & $T_{\rm{eff}}$ [K] & 1000 &  & \\
$f_{\rm{sed}}$ & 3.0  &  &  & log($g$) [cm s$^{-2}$] & 4.75  &  & \\
rfacv & 0.0 &  &  & $[$Fe/H$]$ & 0.0 &  & \\
Age [Gyr] & 3 &  &  & $f$ & 0.5 &  & \\
$M_{\rm{BD}}$ [M$_{\rm{Jup}}$] & 70 &  &  &  &  &  & \\
\cline{1-4}
\multicolumn{4}{c|}{Brock et al. Models} & \\
\cline{1-4}
\multicolumn{2}{l}{\textbf{Day-side Match}} & Yes & No & \multicolumn{2}{l}{\textbf{Model 3}} & Yes & Yes \\
$T_{\rm{int}}$ [K] & 1600 &  &  & $T_{\rm{eff}}$ [K] & 1550 &  & \\
log($g$) [cm s$^{-2}$] & 5.50 &  &  & log($g$) [cm s$^{-2}$] & 5.50 &  & \\
$[$Fe/H$]$ & 0.0 &  &  & $[$Fe/H$]$ & 0.0 &  & \\
$P_{\rm{c}}$ [bar] & 1 &  &  & $P_{\rm{c}}$ [bar] & 20 &  & \\
$K_{\rm{ZZ}}$ & 1.8 &  &  & $K_{\rm{ZZ}}$ & 1.4 &  & \\
$a_{\rm{0}}$ [$\mu$m] & 1.0 &  &  & $a_{\rm{0}}$ [$\mu$m] & 0.3 &  & \\
\multicolumn{2}{l}{\textbf{Night-side Match}} & Yes & No  & \multicolumn{2}{l}{\textbf{Model 4}} & Yes & No \\
$T_{\rm{int}}$ [K] & 1544 &  &  & $T_{\rm{int}}$ [K] & 1544 &  & \\
log($g$) [cm s$^{-2}$] & 5.34 &  &  & log($g$) [cm s$^{-2}$] & 5.34 &  & \\
$[$Fe/H$]$ & 0.0 &  &  & $[$Fe/H$]$ & 0.0 &  & \\
$P_{\rm{c}}$ [bar] & 20 &  &  & $P_{\rm{c}}$ [bar] & 20 &  & \\
$K_{\rm{ZZ}}$ & 1.4 &  &  & $K_{\rm{ZZ}}$ & 1.4 &  & \\
$a_{\rm{0}}$ [$\mu$m] & 0.3 &  &  & $a_{\rm{0}}$ [$\mu$m] & 0.3 &  &
\enddata
\tablecomments{$^*$Calculation for $T_{\rm{eff}}$ of the brown dwarf is described in Section~\ref{sec:LothringerGrid}. When there is no external irradiation, $T_{\rm{int}}$ is the only temperature considered.}
\end{deluxetable*}

\vspace{-0.5 cm}
\subsection{Comparing Day/Night Spectra\\to field Brown Dwarfs}
\label{sec:cloudatlas_disc}
A comparison of NLTT5306B to field brown dwarfs offers an opportunity to test the extent that external irradiation affects emission spectra of irradiated brown dwarfs. Observed brightness variations in field brown dwarfs are mainly driven by heterogeneous clouds in their atmospheres \citep[e.g.,][]{Apai13,Buenzli14,Metchev15,Lew20}. However, in highly irradiated atmospheres like NLTT5306B, variations can be driven by the phase-dependent irradiation and atmospheric circulation. If the day-to-night energy transport is significant on NLTT5306B, this should introduce excess flux into the night-side spectrum.

To test this hypothesis, we searched the Cloud Atlas spectral library of LTY dwarfs \citep[][]{Manjavacas19} for a nearest match using multiple match methods (Section~\ref{sec:cloudatlas}). Both day and night spectral features of NLTT5306B were best matched with the features of a L4.5 brown dwarf, 2MASS~J2224438-015852 \citep[][]{DupuyLiu12, Yang16}, with the night-side being a better fit than the day-side. Silicate clouds are expected to form from condensation in the atmospheres of L dwarfs, but heavy external irradiation might evaporate them. However, with the night-side of NLTT5306B being well-matched with a field brown dwarf spectrum, we conclude that at least within the pressure range probed by our \textit{HST} observations, the night-side of NLTT5306B and the L4.5 both have hemispherically similar thermal and chemical structures.

The key features of field L dwarf spectra, like deep water absorption around 14,000~\AA{}, are present in both day and night spectra of NLTT5306B. The main deviations between the day-side and the field brown dwarfs occurs at wavelengths less than 12,000~\AA{}. Depending on the structure of NLTT5306B's atmosphere, this could be due to the fact that these shorter wavelengths are probing different pressure regions (altitudes) or it may be an effect of irradiation, which we further explore through atmospheric modeling.

\vspace{1 cm}
\subsection{1D Brown Dwarf Atmosphere Models}
\label{sec:compare_models}
To understand the difference between the day- and night-sides of NLTT5306B and identify spectral signatures of irradiation, we created a variety of one-dimensional atmospheric models for both hemispheres. We considered combinations of cloudy vs. cloud-free and non-irradiated vs. irradiated models from three sources: (1) a grid of PHOENIX models (Section~\ref{sec:LothringerGrid}), (2) a grid of PHOENIX models with opacities from \citet{Brock2021} (Section~\ref{sec:BrockGrid}), and (3) a grid of SONORA models (Section~\ref{sec:SonoraGrid}). The best-fit models from each source are listed in Table~\ref{tab:models}.

From all model grids, the night-side spectrum was best matched by atmospheric models that were cloudy and non-irradiated. Given the published system age of at least 5~Gyr, brown dwarf mass of 0.05 M$_{\odot}$, one would expect the internal temperature of a field brown dwarf with the same parameters to be less than 1000~K \citep[based on evolutionary models in][]{Burrows01}. Yet, the night-side of NLTT5306B was best matched by models with internal temperatures between 1500 and 1600~K, suggesting either that NLTT5306B's cooling and evolution was affected by the external irradiation from the white dwarf primary or that the age estimate is too high. We discuss the possible evolutionary scenarios in Section~\ref{sec:evolution}.

In fitting 1D models to the day-side of NLTT5306B, models that included clouds and irradiation were a better fit. A model from the Sonora model grid (Figure~\ref{fig:sonoramodels}) provided the closest match to the extracted spectrum, with a smaller $f_{\rm{sed}}$ and higher rfacv than the night-side model. However, no fully satisfactory model fits could be found to match spectral features at all wavelengths simultaneously, such as the contrast between the flux at 1.3~$\mu$m and the water absorption at 1.4~$\mu$m and the slopes of the spectrum in the $J$- and $H'$-bands.

Given that NLTT5306B is tidally-locked in a tight orbit around a WD primary, it was perhaps not surprising that the day-side spectrum best matched with an irradiated model. This was not necessarily in contrast to the atmospheric parameters found in Section~\ref{sec:longtempdist}. Even with a strong $f_{\rm{irr-red}}$ = 0.70$\pm$0.02, the ratio of irradiated flux to internal heat flux for NLTT5306B is small with respect to other WD+BD systems (see Section~\ref{sec:other_wdbds}). This relatively strong internal heat flux would then still dominate on the night side, whereas the day side is constantly being irradiated, leading to irradiated models being better fits.

\vspace{-0.1 cm}
\subsection{Comparison to Hot Jupiters}
Irradiated brown dwarfs differ from field brown dwarfs in that they experience significant external irradiation from a host star, transforming them into excellent analogs to hot Jupiters. One of the key mechanisms believed to impact hot Jupiter atmospheres is the formation/disruption of clouds, which strongly affect atmospheric opacities and pressure-temperature profiles \citep[][]{Burrows97, Marley99, Ackerman01, Burrows01, Lee16, HengDemory13}. Observations of irradiated brown dwarfs gives us ideal templates when searching for patterns between hot Jupiter properties and their cloud properties.

NLTT5306B orbits its primary star on a much closer orbit than most hot Jupiters. However, the relatively small size of NLTT5306A means the emitting surface area is also small, resulting in an equilibrium temperature of NLTT5306B on the cooler end of temperatures encompassed by hot and ultra-hot Jupiters (e.g. from $\sim 500 - 5000$ K). For hot Jupiters, the size and intensity of equatorial jets are believed to scale with day-side irradiation \citep[][]{ShowmanGuillot02, Showman20}. In the cloudless GCM of NLTT5306B, a thin jet is present at the equator, with retrograde jets at adjacent latitudes due to the fast rotation (Figure~\ref{fig:cloudfree_gcm}). A more detailed GCM is needed to confirm the characteristics of the equatorial jet in this system, but other simulations of irradiated brown dwarfs \citep[][]{TanShowman20_rotationWDBDs, Lee20} suggest that the rapid rotational timescale (few hours) in WD+BD binaries causes the narrow equatorial jet.

\begin{figure}
\begin{center}
\includegraphics[width=0.47\textwidth]{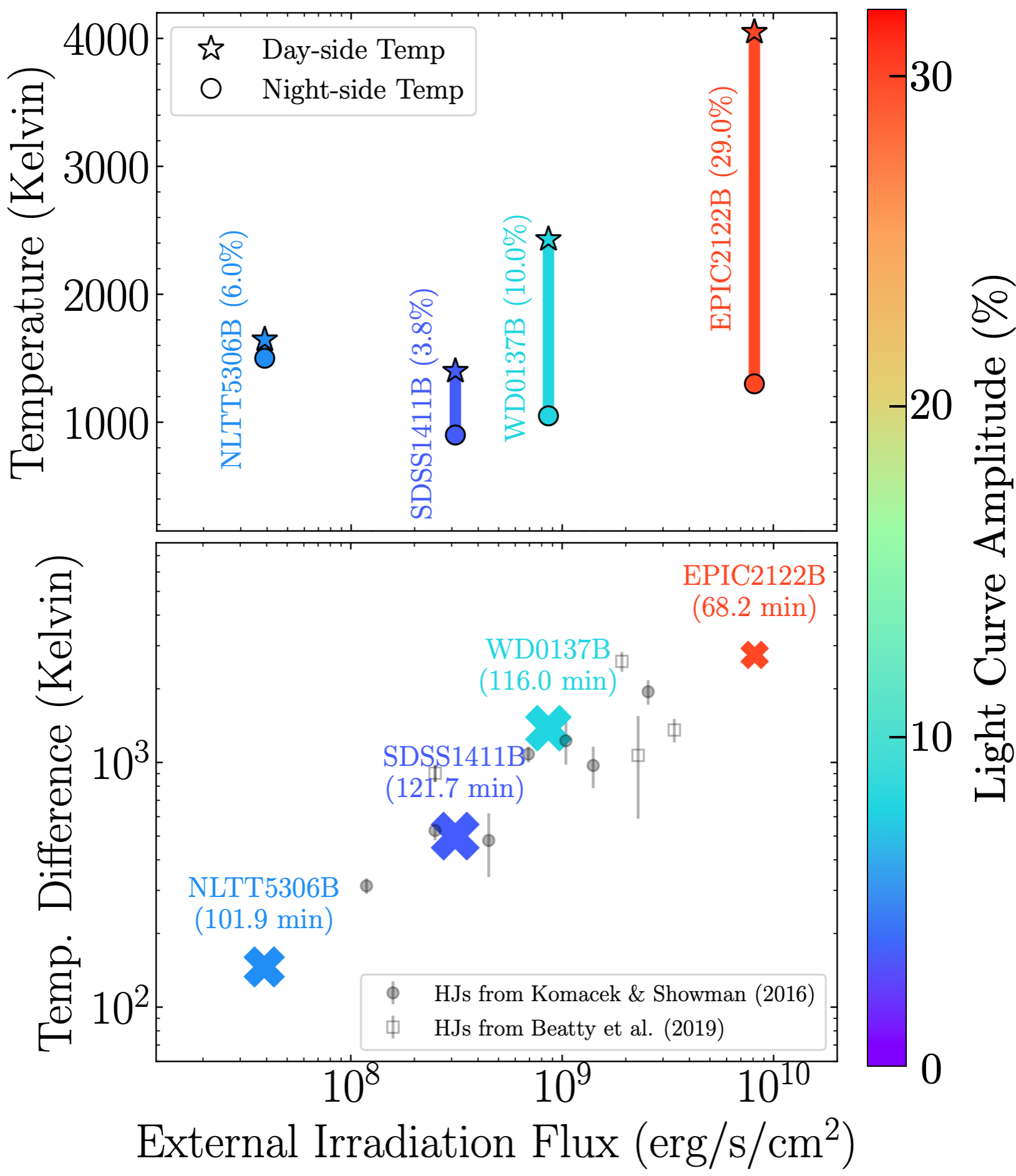}
\caption{\textbf{Top:} Temperature ranges of NLTT5306B, SDSS1411B, WD0137B, and EPIC2122B versus external irradiation flux from their respective primary white dwarfs. Day- (star) and night-side (circle) hemisphere-averaged brightness temperatures are shown (NLTT5306B has been corrected for inclination). Color represents amplitude of broadband light curves, also labeled next to each name. NLTT5306B does not appear to follow the trend set by the other three systems, where we would expect NLTT5306B to be colder with a smaller light curve amplitude than SDSS1411B. \textbf{Bottom:} Day-Night temperature difference compared to external irradiation flux. Hot Jupiters from \citet{KomacekShowman16} and \citep[][]{Beatty19}. Sizes of WD+BD markers are scaled according to system period, also labeled under each name. A direct relationship between Day-Night difference and irradiation can be seen here despite NLTT5306B's deviation in the panel above, but there is no obvious relation with system period.}
\label{fig:compareWDBDs}
\end{center}
\end{figure}

The bottom panel of Figure~\ref{fig:compareWDBDs} shows a comparison of Day-Night temperature contrast to external irradiation flux for NLTT5306B, three other WD+BD systems, and hot Jupiters from \citet{KomacekShowman16}. The hot Jupiters included are HD 189733b  \citep[][]{Knutson09, Knutson12}, HD 209458b \citep[][]{Crossfield12, Zellem14}, HD 149026b \citep[][]{Knutson09}, WASP-14b \citep[][]{Wong15}, WASP-19b \citep[][]{Wong16}, HAT-P-7b \citep[][]{Wong16}, and WASP-12b \citep[][]{Cowan12}. We also include additional hot Jupiters from \citet{Beatty19} that are not present in the \citet{KomacekShowman16} sample. These hot Jupiters are WASP-18b \citep[][]{Maxted13}, WASP-33b \citep[][]{Zhang18, Chakrabarty19}, WASP-43b \citep[][]{Stevenson17}, WASP-103b \citep[][]{Kreidberg18}. A direct relationship between temperature contrast and external irradiation flux can be seen for both the WD+BD systems and the hot Jupiters. This trend was predicted for hot Jupiters by \citet{ShowmanGuillot02}, further supporting the assumption that irradiated brown dwarfs are excellent analogs for hot Jupiters.

One of the few hot Jupiters with a measured spectroscopic phase curve is WASP-43b \citep[][]{Stevenson14}, which has an equilibrium temperature around 1426~K \citep[][]{Esposito17} and a primary host star of spectral type K7 \citep[][]{Salz15}. From the spectroscopic phase curve of WASP-43b, \citet{Stevenson14} observed wavelength-dependent phase offsets, much like what is seen in NLTT5306B. In WASP-43b, there was a visible correlation between the day-side thermal emission contribution levels and phase-curve peak offsets as a function of wavelength \citep[see Figure~S7 in][]{Stevenson14}. This is not seen in NLTT5306B when using the best-fit models, but is likely due to our probed pressures not changing by much across our observed \textit{HST} wavelengths.

\subsection{Comparison to WD~0137B,\\ EPIC~2122B, and SDSS1411-B}
\label{sec:other_wdbds}

NLTT5306B is the fourth irradiated brown dwarf to be studied in the ``Dancing with the Dwarfs'' data set, obtained under the \textit{HST} program GO-15947 (PI: Apai). WD~0137B, EPIC~2122B \citep[][]{Zhou22}, and SDSS1411-B \citep[][]{Lew22} are the previous three that have already been analyzed and published. Analysis of additional data sets in this program are ongoing, but here we present a preliminary comparison with these four WD+BD systems (Figure~\ref{fig:compareWDBDs}).

WD~0137B, EPIC~2122B, and SDSS1411-B all receive higher levels of irradiation\footnote{Using $T_{\rm{eff,WD}}=16,500$~K, $a=0.65$~R$_{\odot}$, $R_{\rm{WD}}=0.0186$~R$_{\odot}$ for WD~0137B, $T_{\rm{eff,WD}}=24,900$~K, $a=0.44$~R$_{\odot}$, $R_{\rm{WD}}=0.017$~R$_{\odot}$ for EPIC~2122B, and $T_{\rm{eff,WD}}=13,000$~K, $a=0.003$~AU, $R_{\rm{WD}}=0.0179$~R$_{\odot}$ for SDSS1411-B.} from their primary white dwarfs than NLTT5306B (approximately 22, 208, and 8 times higher, respectively). They also exhibit differing light curve amplitudes (i.e., day/night-side intensity differences) of 10, 29, and 3.8\%, respectively \citep[][]{Zhou22,Lew22}. In the top panel of Figure~\ref{fig:compareWDBDs}, following the relationship established by the other three WD+BD systems, we would expect NLTT5306B to have cooler brightness temperatures and a smaller light curve amplitude than SDSS1411-B. Yet, NLTT5306B deviated from this expectation, which we attribute to the high internal heat flux.

Conversely, in the bottom panel of Figure~\ref{fig:compareWDBDs}, NLTT5306B appeared to follow a direct relationship between external irradiation and day/night temperature contrast. Note that the temperature difference here was corrected for the effect of inclination (see Section~\ref{sec:longtempdist}). After applying a $10\%$ flux increase (decrease) to the day (night) BD spectra and re-deriving brightness temperatures, the temperature difference increased from $93\pm32$~K (Figure~\ref{fig:brighttemps}) to $146\pm33$~K. Another important factor in day/night temperature contrast is the rotation rate of the BD \citep[][]{TanShowman20_rotationWDBDs}, which is inversely proportional to the orbital period in tidally locked companions. Yet NLTT5306B has the second shortest period, and thus second highest rotation rate, in this sample. \citet{TanShowman20_rotationWDBDs} predicted that day/night temperature contrasts would increase with increasing rotation rate, however NLTT5306B deviates from this expectation, another consequence of the strong internal heat flux. So far, it is clear that this growing sample of WD+BD pairs is diverse in many aspects, including irradiation strength, rotation rate, and day/night temperature contrast.

Additionally, in the wavelengths observed by \textit{HST}, the day- and night-sides of WD~0137B, EPIC~2122B, and SDSS1411-B show significantly different spectral features due to irradiation (see Figure 8 in \cite{Zhou22} and Figure 9 in \cite{Lew22}), whereas NLTT5306B displays very similar features between both hemispheres. This is not unexpected, as NLTT5306B receives much weaker irradiation than the other WD+BD systems. This may indicate that higher irradiation induces greater contrasts, but other factors -- such as heat redistribution efficiency and opacities -- also affect this relationship.

A key difference between NLTT5306B and the other three WD+BD systems is the presence of wavelength-dependent phase offsets. The complex nature of the relative phase offsets seen in Figures~\ref{fig:offs_amps_k1} and \ref{fig:offs_amps_k2} were not present in the phase curves WD~0137B, EPIC~2122B, or SDSS1411-B. More advanced, self-consistent 3D modeling across all systems in this program will be required to confidently identify the source of this difference.

\begin{figure}
\begin{center}
\includegraphics[width=0.45\textwidth]{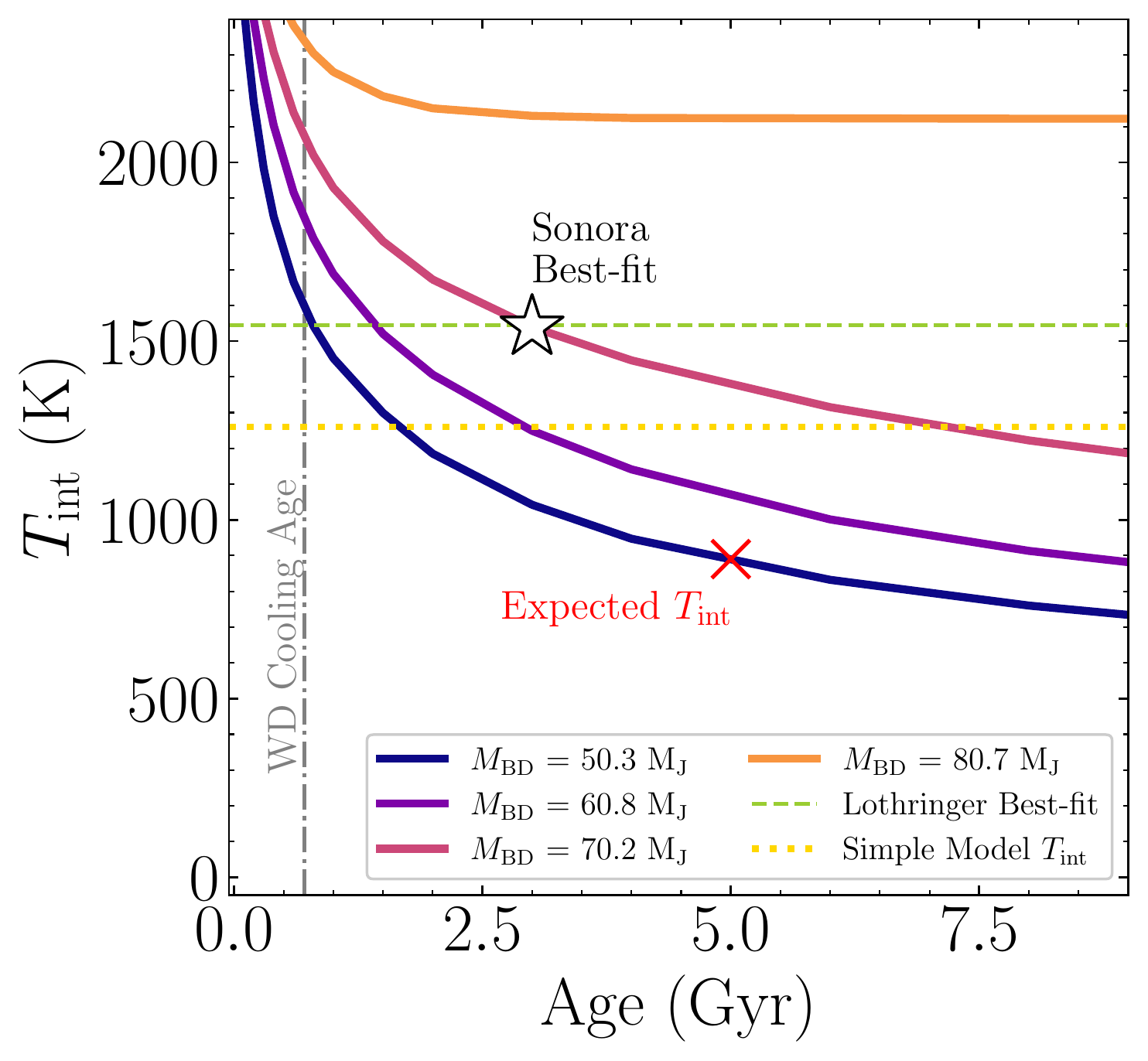}
\caption{Evolution tracks of four brown dwarf masses according to Sonora Bobcat 2018 tables. A gray dot-dashed line depicts the published WD cooling age from \citet{Steele13}. A red X marks the predicted temperature for a $M_{\rm{BD}} \sim 50$ M$_{\rm{J}}$ brown dwarf at 5 Gyr. Observations and models of NLTT5306B do not fit to this prediction. The best-fit night-side $T_{\rm{int}}$ from the Lothringer model grid is shown as a green dashed line. Predicted $T_{\rm{int}}$ from the simple model grid is shown as a golden dotted line. The best night-side match from the Sonora model grid was a $M_{\rm{BD}}=70.2$ M$_{\rm{J}}$ at 3 Gyr.}
\label{fig:evolution}
\end{center}
\end{figure}

\vspace{0.5 cm}
\subsection{Evolution of NLTT5306B}
\label{sec:evolution}
Brown dwarfs cool as they age, with more massive brown dwarfs cooling at a slower rate than their less massive counterparts. This is illustrated in Figure~\ref{fig:evolution}, where the cooling curves for four brown dwarf masses from the Sonora Bobcat 2018 tables \citep[][]{Marley18} are shown. \citet{Steele13} estimated the system age of NLTT5306 to be at least 5~Gyr, based on the systemic WD velocity, particularly the $U$ component \citep{JohnsonSoderblom87}, which generally placed NLTT3506 within the thick-disc population (see Figure 4 of \cite{Pauli06}). Based on Figure~\ref{fig:evolution}, if NLTT5306B were a 5~Gyr, 56$\pm$3 M$_{\rm{Jup}}$ field brown dwarf, one would expect the interior heat to be less than 1000~K (presented as a red X). However, our derived brightness temperatures and simple model grid results suggested that NLTT5306B has internal temperature around 1200$-$1300~K.


It is clear that the $T_{\rm{int}}$ evolution of NLTT5306B does not fit to what is expected for a $\sim$50 M$_{\rm{Jup}}$ field brown dwarf at 5~Gyr, suggesting either that (1) interaction with the white dwarf primary has affected the evolutionary cooling of the brown dwarf or (2) the system is actually younger than 5~Gyr. Two explanations for scenario (1) could include a reduced internal cooling rate \citep[e.g.,][]{Burrows07, Chabrier07, Chabrier_Baraffe07, Fortney21}, a deep internal deposition of heat from the primary star \citep[][]{Guillot_Showman02, Batygin_Stevenson10, Sainsbury-Martinez21}. However, the high internal heat flux in NLTT5306B suggests that the cooling rate has not been sufficiently reduced at the moment. Furthermore, even if 100\% of the stellar irradiation was deposited into the interior of the BD, it would still be a small fraction of the observed internal flux, meaning that both mechanisms for scenario (1) are unlikely.

Scenario (2) is consistent with the high night-side temperature. Previously, \citet{Steele13} estimated the systemic velocities for NLTT5306 using a distance of $D$=71$\pm$4 pc, equatorial coordinates and parallaxes, and radial velocities from their XSHOOTER and Hobby-Eberly Telescope (HET) data. Focusing on the $U$ value, $U\approx70$~km~s$^{-1}$, they concluded that NLTT5306A was most likely a thick-disc object, based on Figure~4 from \citet{Pauli06}. With updated, higher precision parameters from Gaia DR3\footnote{https://gea.esac.esa.int/archive/documentation/GDR3/} \citep[][]{GaiaCollaboration2016}, kinematics for NLTT5306 and over 3,000 other white dwarfs were calculated by \citet{Raddi22}.

To determine the new age of NLTT5306B, we downloaded Tables 4 and 7 from \citet{Raddi22} and extracted the ages of thin-disc only systems, i.e. only systems with $e<$ 0.27. We then calculated two total ages for every system by summing the WD cooling age with the two progenitor age estimates from the \citet{Catalan08} (C08) and \citet{Cummings18} (C18) initial-to-final-mass relations. Since NLTT3506A has a cooling age of 710 Myr, we ignored ages younger than 0.7 Gyr, resulting in age peaks around 1 Gyr for C08 and 2 Gyr for C18. Therefore, we approximate the system age of NLTT5306B to most likely be 1-2 Gyr, which essentially removed the age vs. internal temperature discrepancy for NLTT5306B and reaffirmed our conclusion that this BD is dominated by internal heat flux.

\section{Conclusions} \label{sec:conclusions}

In this study, we presented time-resolved, high-precision spectrophotometric, rotational phase mapping of the irradiated brown dwarf NLTT5306. The key findings of our study are as follows:

\begin{itemize}
    \item We presented high-quality \textit{HST}/WFC3/G141 phase-resolved spectra of white dwarf-brown dwarf binary NLTT5306. The observations sampled approximately 90\% of the rotational phase of this system and exhibited moderate photometric and spectroscopic variations.
    
    \item We modeled the light curve of NLTT5306 to within 0.5\%. The period of the favored 1st-order phase curve model created in this work (102.05$\pm$0.07 min) is longer than the published period of 101.88$\pm$0.02 min \citep[][]{Steele13}.

    \item The synthetic light curves made in the $J$-, Water Absorption, and $H'$-bands had peak-to-trough amplitudes between $\sim$5-7\% and relative phase offsets up to 6.8$\pm$0.3$^\circ$. 
    
    \item Synthetic light curves made from narrower wavelength bins (Section~\ref{sec:offs_amps}) exhibited a prominent wavelength dependence on amplitudes and phase offsets. This indicated a complex longitudinal-vertical atmospheric structure.
    
    \item After modeling and subtracting the white dwarf contribution, we extracted phase-resolved brown dwarf spectra, with a day-side approximately 40\% brighter than the night-side. Both the day and night spectra were well fit by an L4.5 spectral type field brown dwarf spectrum from the Cloud Atlas catalog.
    
    \item Although the day and night brightness temperatures of NLTT5306B were highly wavelength dependent, the temperature difference was nearly constant at 93~K (or 5\% of the average day side temperature). This suggested efficient day/night heat redistribution.
    
    \item A simple radiative and energy redistribution model reproduced well both the observed day- and night-side temperatures. The model suggested an $A_B$ = 0.28$\pm$0.07, $f_{\rm{irr-red}}$ = 0.70$\pm$0.02, $T_{\rm{non-irr}}$ = 1260$\pm$20 K, and $i$ = $78.0^{+1.1}_{-7.6}$ degrees for NLTT5306B.
    
    \item The day-side of NLTT5306B was well but not perfectly fit with an irradiated, cloudy 1D model. The night-side was fit by a non-irradiated, cloudy model.
    
    \item A general circulation model is presented that explains the observed, relatively small day-to-night-side temperature difference. The small difference is due to the relatively weak irradiation (with respect to the internal heat flux).
    
    \item The internal temperatures predicted for NLTT5306B by Sonora Bobcat 2018 and Lothringer models were inconsistent with the age previously reported \citep[][]{Steele13}. Our reassessment of the age estimate suggested NLTT5306A to be much younger ($\sim$1$-$2~Gyr), which was fully consistent with the evolutionary model predictions.

    \item The range of likely inclinations for NLTT5306 were constrained to $i = 78.0^{+1.1}_{-7.6}$ degrees, yielding a flux correction factor of $+10\%$ for the day-side and $-10\%$ for the night-side. We explored the impact this uncertainty would have on the results of simple atmosphere model, and found that only one parameter, Bond albedo, was sensitive to the unknown inclination.
\end{itemize}

Our comprehensive analysis shows that the brown dwarf in the NLTT5306 system is one of the least irradiated brown dwarfs known to be in a tidally-locked orbit, bridging the gap between highly irradiated brown dwarfs and field brown dwarfs. Given the re-assessed age, this is also a relatively young system where the internal heat still outcompetes the irradiation from its white dwarf host.

\begin{acknowledgments}
This material is based upon work supported by the National Science Foundation Graduate Research Fellowship under Grant No. DGE-1746060. Any opinion, findings, and conclusions or recommendations expressed in this material are those of the authors(s) and do not necessarily reflect the views of the National Science Foundation. \textit{HST} data presented in this paper were obtained from the Mikulski Archive for Space Telescopes (MAST) at the Space Telescope Science Institute. 
The specific observations analyzed can be accessed via \dataset[10.17909/hmqg-zg25]{https://doi.org/10.17909/hmqg-zg25}. Support for Program number HST-GO-15947 was provided by NASA through a grant from the Space Telescope Science Institute, which is operated by the Association of Universities for Research in Astronomy, Incorporated, under NASA contract NAS5-26555. This publication makes use of data products from the Two Micron All Sky Survey, which is a joint project of the University of Massachusetts and the Infrared Processing and Analysis Center/California Institute of Technology, funded by the National Aeronautics and Space Administration and the National Science Foundation. This work has made use of data from the European Space Agency (ESA) mission {\it Gaia} (\url{https://www.cosmos.esa.int/gaia}), processed by the {\it Gaia} Data Processing and Analysis Consortium (DPAC,
\url{https://www.cosmos.esa.int/web/gaia/dpac/consortium}). Funding for the DPAC has been provided by national institutions, in particular the institutions participating in the {\it Gaia} Multilateral Agreement.

\end{acknowledgments}

\facility{Hubble Space Telescope (WFC3)}

\software{Astropy \citep{astropy2013,astropy2018},  
          Source Extractor \citep[][]{Bertin96_sextractor},
          Numpy \citep[][]{numpy2020},
          Scipy \citep[][]{scipy2020-NMeth},
          Matplotlib \citep[][]{matplotlib},
          HSTaXe (\url{https://github.com/spacetelescope/hstaxe})
          }

\appendix

\section{Wavelength-Dependent Phase Offsets and Modulation Amplitudes}
\label{sec:offs_amps}

Motivated by the $>$5 degree phase offsets and $>$1\% amplitude differences between the $H'$-band and the other synthetic light curves, we performed a more detailed analysis to pinpoint which wavelengths contributed most to the observed behavior. By narrowing down the wavelengths, we can also point to the chemical species that is the dominant absorber in that wavelength range, helping us a paint a clearer picture of the atmospheric dynamics at play.

\begin{figure}
\begin{center}
\includegraphics[width=0.69\textwidth]{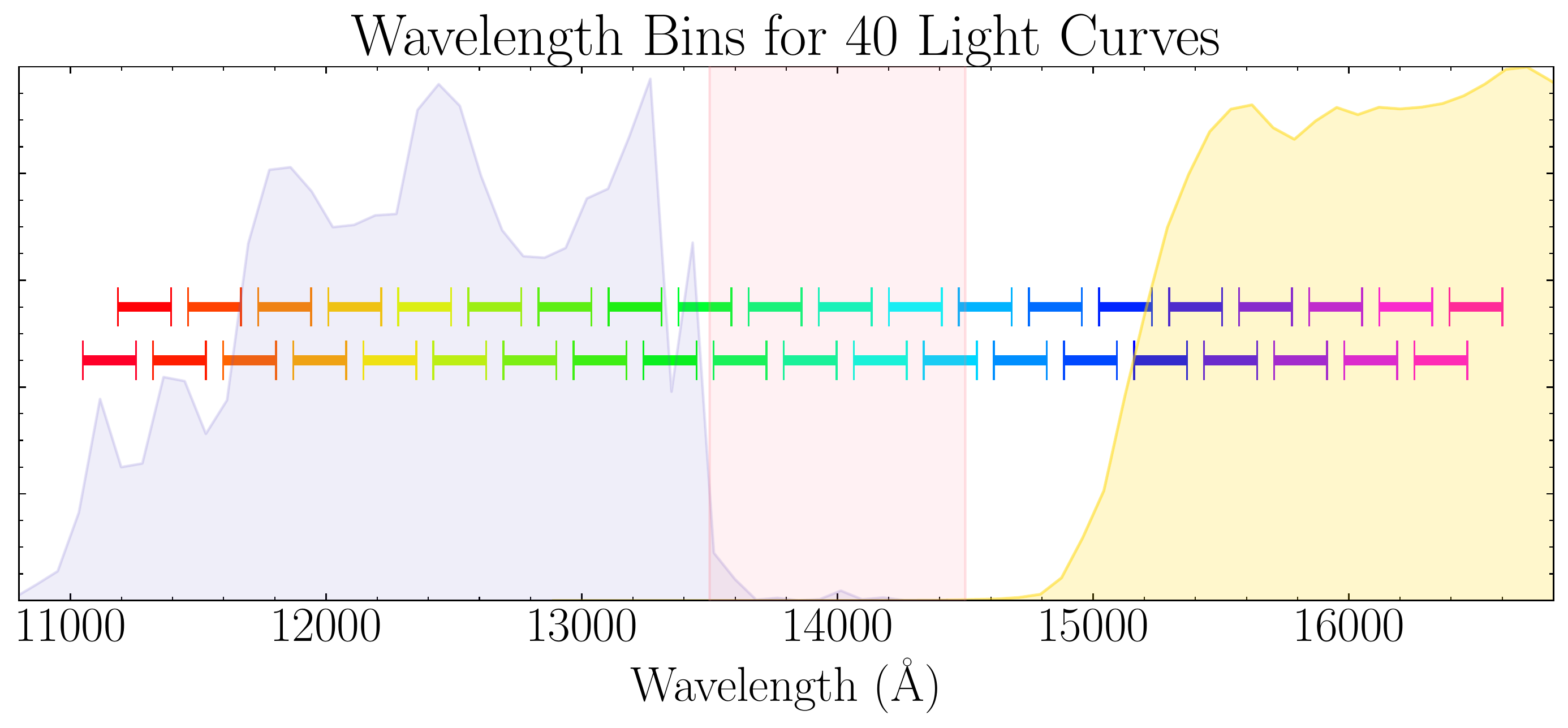}
\caption{Locations of new wavelength bins used to investigate the wavelength dependence of phase offsets and amplitudes. The old light curve filters, 2MASS $J$- (purple), Water (pink), and 2MASS $H$-band (yellow), are shown as a reference, but are not applied to the new light curves. For easier viewing, the new wavelength bins have been staggered on top of each other. Each new bin spans 137~\AA{} with $\sim$2/3 of each bin overlapping with its neighboring bins.}
\label{fig:narrowLC_showbins}
\end{center}
\end{figure}

\begin{figure}
\begin{center}
\includegraphics[width=0.99\textwidth]{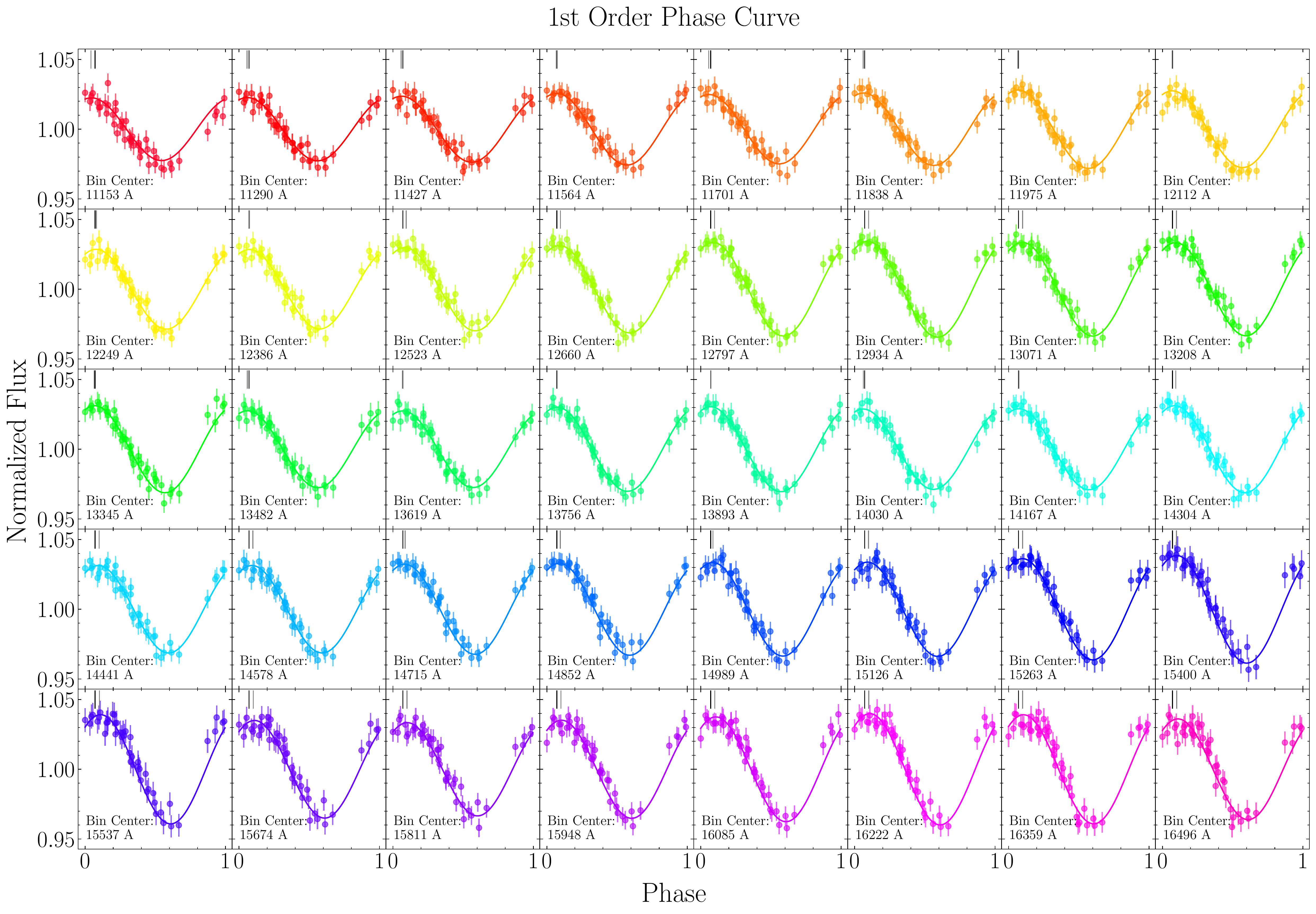}
\caption{Period-folded phase curves created from 40 narrow (136~\AA{}) wavelength bins and assuming a 1st-Order phase curve model. The wavelength of the bin centers are noted in the bottom left of each plot. Phase offsets for these phase curves are calculated from the difference between the peaks of these models (gray vertical line) and the $J$-band peak (black vertical line). So, the farther apart the vertical lines appear in phase space, the greater the phase offset.}
\label{fig:narrowLC_k1}
\end{center}
\end{figure}

\begin{figure}
\begin{center}
\includegraphics[width=0.99\textwidth]{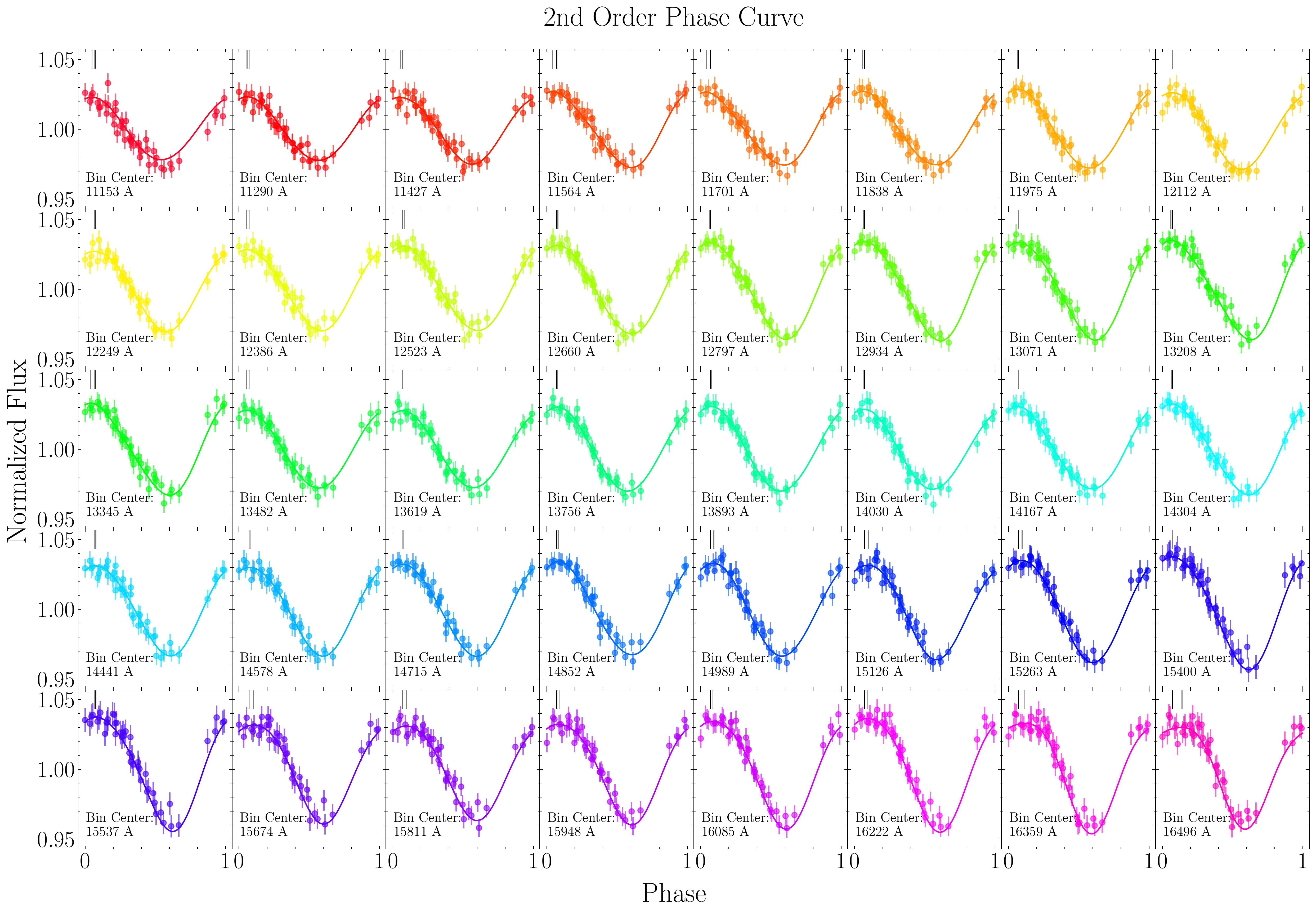}
\caption{Similar to Figure~\ref{fig:narrowLC_k1}, except these phase curves models are made with a combination of $k$=1,2 sinusoidal functions (see Equation~\ref{eq:combo_sines}).}
\label{fig:narrowLC_k2}
\end{center}
\end{figure}

We split the spectra into 40 wavelength bins that are each 137~\AA{} wide, allowing for some overlap between the new bins, as shown in Figure~\ref{fig:narrowLC_showbins}. When creating the new light curves, filter response functions were not applied, i.e. 2MASS $J$ and $H'$ used previously for the synthetic light curves. Our intention was to ensure that the differences between bins were real and not the result of being exaggerated by the tail ends of response functions. The 40 new light curves are presented in Figures~\ref{fig:narrowLC_k1} and \ref{fig:narrowLC_k2}, along with their 1st and 2nd order phase curves models, respectively.

\begin{figure}
\begin{center}
\includegraphics[width=0.59\textwidth]{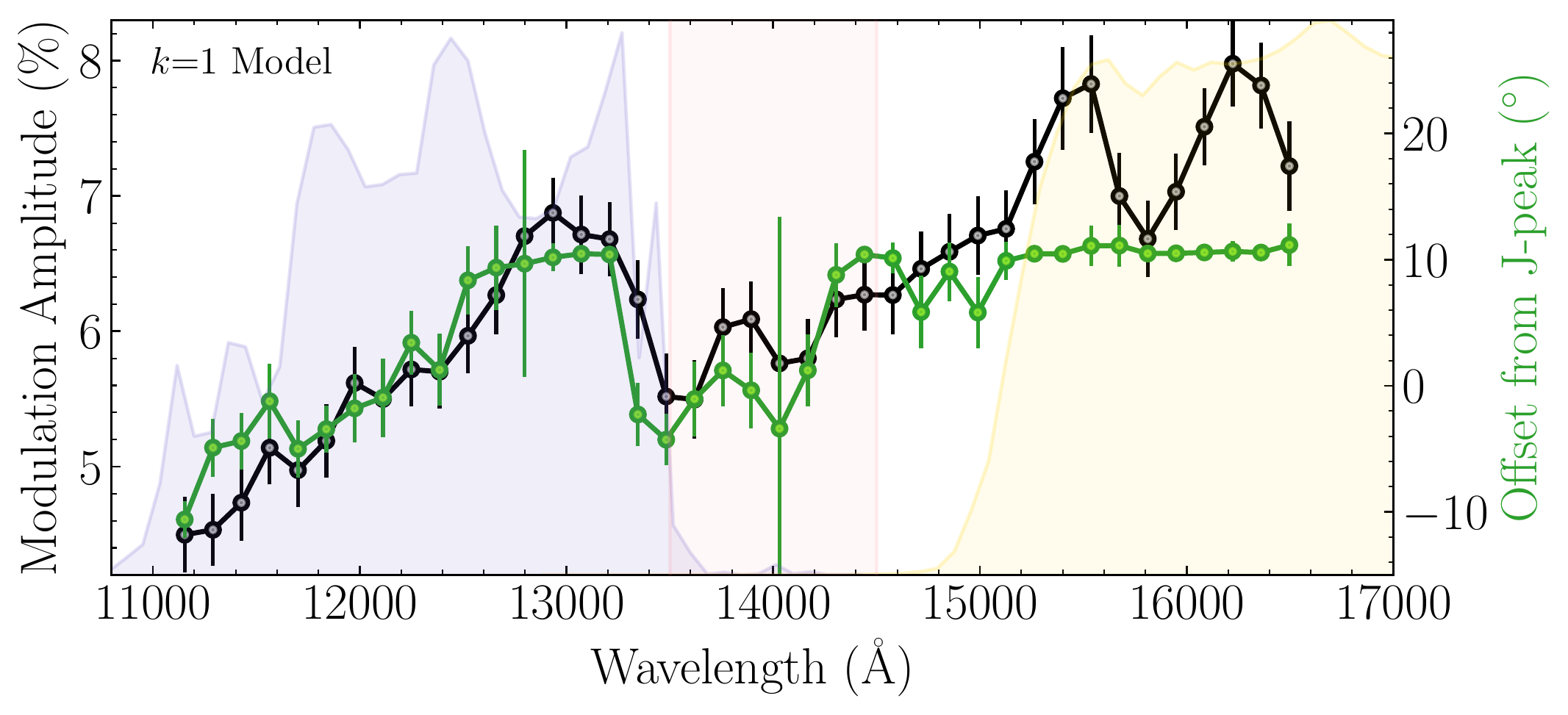}
\caption{The modulation amplitudes (black) and phase offsets (green) of the 1st-order ($k$=1) phase curves, corresponding to Figure~\ref{fig:narrowLC_k1}. The high SNR of our data allows us to conduct this detailed investigation into the wavelength dependence on amplitude and phase. Although they were not applied here, the old light curve filters (i.e. 2MASS $J$, Water Absorption, and 2MASS $H$) are shown in the background as a reference. Error bars depict 1$\sigma$ uncertainty from the MCMC chains. The phase offset curve seems to generally follow the same shape as the amplitude curve until reaching the $H'$-band wavelengths. After 15,000~\AA{}, the phase offsets appear to remain constant, even though the amplitude curve reaches its highest peaks here.}
\label{fig:offs_amps_k1}
\end{center}
\end{figure}

\begin{figure}
\begin{center}
\includegraphics[width=0.59\textwidth]{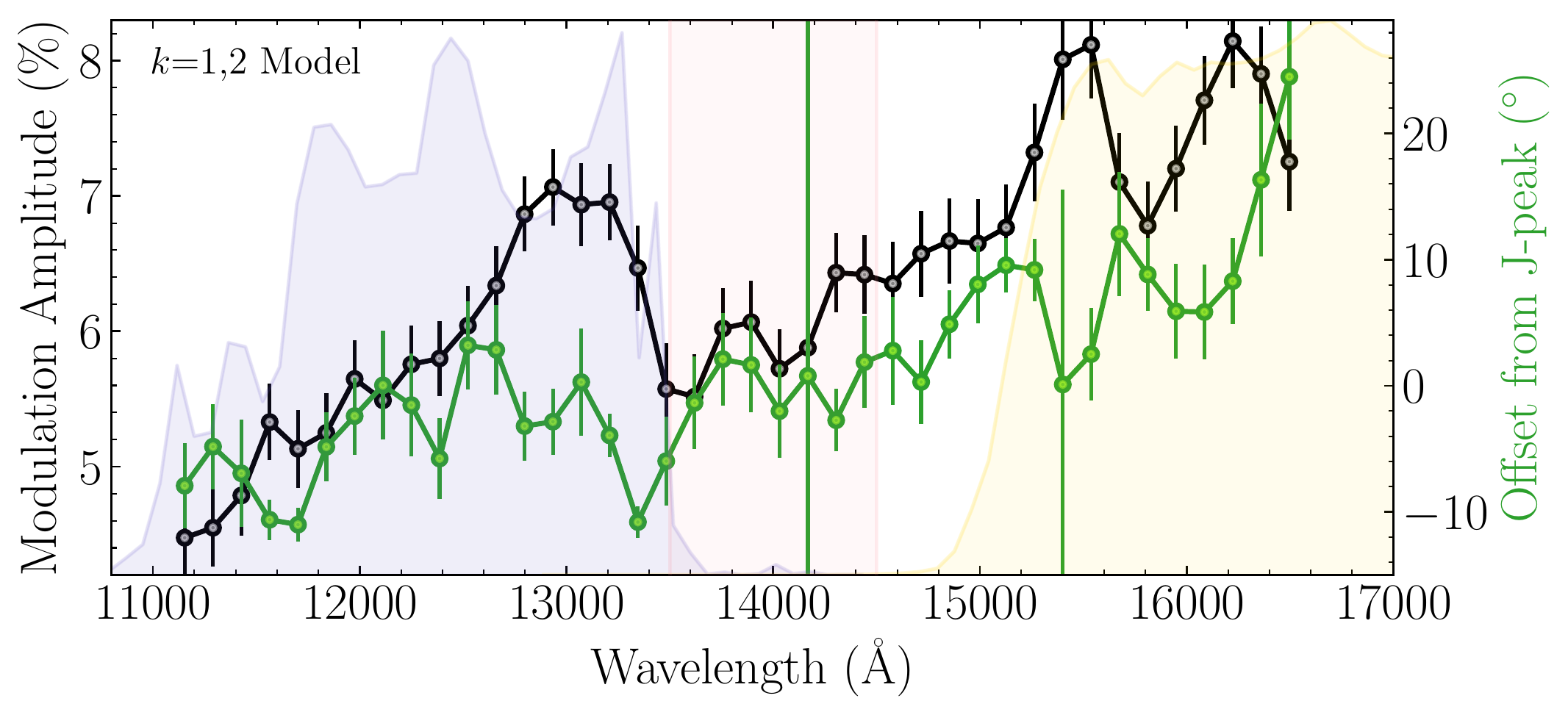}
\caption{Similar to Figure~\ref{fig:offs_amps_k1}, except these amplitudes and offsets correspond the 2nd order ($k$=1,2) phase curve models in Figure~\ref{fig:narrowLC_k2}. Here, the phase offset curve again seems to follow to the modulation amplitude curve, except for when the amplitude curve peaks above 7\% around three wavelengths: 12,900, 15,450 and 16,250~\AA{}. When the modulation amplitude curve begins to peak, the phase offset curve begins to dip. It remains unknown why an anti-correlation would be present.}
\label{fig:offs_amps_k2}
\end{center}
\end{figure}

Similar to the phase curve fitting functions described before, we used an MCMC to find the best-fit phase curve for both a 1st and 2nd order sinusoidal function. The residuals produced by subtracting each model from the data didn't show much preference for either order, so we included both for comparison. For each light curve, we produced 10,000 phase curve models with a burn of 1,000, leaving 9,000 models to find the best-fit phase curve parameters. The resulting median and 1-sigma uncertainties for phase offsets and modulation amplitudes are shown in Figures~\ref{fig:offs_amps_k1} and \ref{fig:offs_amps_k2}. Phase offsets are calculated with respect to the peak of the $J$-band phase curve model.

\restartappendixnumbering

\section{Flux Scaling Cloud Atlas Spectra to Photometry of NLTT5306B}
\label{sec:fluxscale_cloudtlas}
In Table~1 of \citet{Manjavacas19}, the authors report a 2MASS $J$-band magnitude for each L, T, Y dwarf. We use this magnitude to photometrically scale the spectra to the expected $J$-band magnitudes of the day- and night-sides of NLTT5306B. We estimated what the day- and night-side magnitudes would be by solving the equation for apparent magnitudes and flux ratios:
\begin{equation}
    m_{J,\rm{BD}} = m_{J, \rm{WD+BD}} - 2.5~\rm{log_{10}}\Big( \frac{\overline{F}_{J\rm{,BD}}}{\overline{F}_{J\rm{,WD+BD}}}  \Big)
\end{equation}

where $m_{J\rm{,BD}}$ is the brown dwarf magnitude we are looking for, $m_{J\rm{,WD+BD}}$ is the $J$-band magnitude of the white dwarf + brown dwarf system ($m_{J\rm{,WD+BD}}$=16.23),  $\overline{\rm{F}}_{J\rm{,WD+BD}}$ is the mean flux value in the $J$-band of the observed white dwarf + brown dwarf spectrum, and $\overline{\rm{F}}_{J\rm{,BD}}$ is the average flux in the $J$-band of the extracted brown dwarf spectrum. From this equation, we found that $m_{J}$(BD$_{\rm{Day}}$) = 18.08 $\pm$ 0.22 and $m_{J}$(BD$_{\rm{Night}}$) = 18.40 $\pm$ 0.31. When using the ``flux scaling'' match method (e.g., top row of Figure~\ref{fig:cloudatlas}), the Cloud Atlas spectra were scaled to these magnitudes.

\section{Grid Search Among Bond Albedo, Irradiation Redistribution Fraction, Non-Irradiated BD Temperature, and Inclination}
\label{sec:tempmap_gridsearch}

We conducted a grid among 4 parameters: (1) Bond albedo, $A_B$ = [0.0, 1.0] in increments of 0.025, (2) irradiation redistribution fraction, $f_{\rm{irr-red}}$, sampled between 0.0 and 1.0 in increments of 0.02, (3) non-irradiated brown dwarf temperature, $T_{\rm{non-irr}}$, sampled from 800 to 2000~K in increments of 10~K, and (4) inclination, $i$, from 57 to 85 degrees in increments of 0.5. For each combination of parameters, the first step was to apply a uniform day- and night-side temperature set by the input $T_{\rm{non-irr}}$. Then, incident radiation from the WD was added to the day-side, and the value of $A_B$ determined how much was reflected away. In this tidally-locked set-up, the substellar point is the hottest location on the BD surface for all $A_B<$1.0. Next, we redistributed a fraction of the incident radiation from the day-side, determined by $f_{\rm{irr-red}}$, and evenly distributed it onto the BD entire surface. If $f_{\rm{irr-red}}$ = 1.0, all irradiated flux would be evenly redistributed, resulting in a uniform temperature map. Finally, we added the effect of inclination by taking a number of rows at the top of the temperature map and switching them between day and night hemispheres. The number of rows were determined by the inclination parameter, where a lower inclination would result in a larger amount of rows, due to the greater percentage of the opposite hemisphere being seen. To determine the best-fit set of parameters, we calculated residuals between each temperature distribution map and our observables, i.e. brightness temperatures.

Since brightness temperatures are a hemisphere-integrated quantity, we calculated the day and night hemisphere-integrated temperatures for each map ($T_{\rm{map,day}}$ and $T_{\rm{map,night}}$), where day is defined as $-$90$^{\circ}>$longitude$>$90$^{\circ}$ and night $-$90$^{\circ}\geq$longitude$\leq$90$^{\circ}$ (see Figure~\ref{fig:radequilmap}). From our brightness temperatures and 1D modeling results, we assumed the observed night-side temperature to be $T_{\rm{BD,night}}$=1500$\pm$20 K. Then, from Figure~\ref{fig:brighttemps}, we assumed $T_{\rm{BD,day}}$=$T_{\rm{BD,night}} + 93$ K=1593$\pm$30 K. The residuals were calculated via the following:
\begin{equation}
    R = \left( \frac{T_{\rm{BD,day}} - T_{\rm{map,day}}}{\sigma_{\rm{BD,day}}} \right)^2 + \left( \frac{T_{\rm{BD,night}} - T_{\rm{map,night}}}{\sigma_{\rm{BD,night}}} \right)^2
\end{equation}

We accepted the smallest value of R as the correct answer, and also checked that the combination of parameters was within the realm of realistically physical values. The combination that resulted in the smallest R was $A_B$ = 0.28$\pm$0.07, $f_{\rm{irr-red}}$ = 0.70$\pm$0.02, $T_{\rm{non-irr}}$ = 1260$\pm$20 K. and $i = 78.0^{+1.1}_{-7.6}$ degrees. While this temperature distribution model is fairly simple and cannot capture more complex dynamics, we believe such an approach is a useful starting point for understanding observations of tidally-locked atmospheres.

\section{Effect of Inclination}
\label{sec:inclinations}

\begin{figure}
\begin{center}
\includegraphics[width=0.95\textwidth]{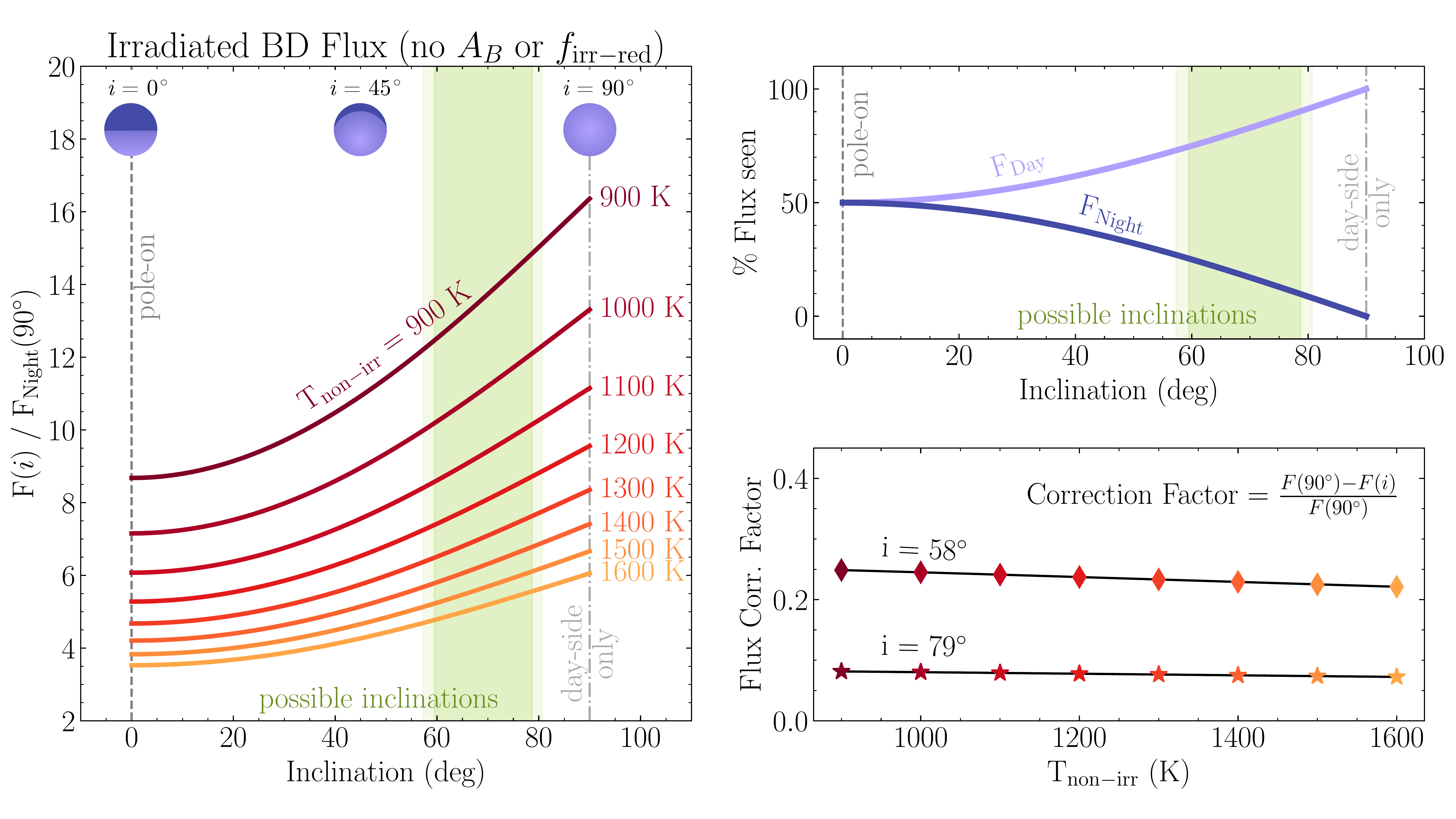}
\caption{\textbf{ Left:} The effect of inclination on derived flux seen at an orientation where the day-side dominates, .i.e, the brown dwarf is farthest from the observer. Inclinations other than $i = 90^\circ$ result in the night-side being partially visible, which decreases the derived flux. The green shaded region shows the possible inclinations for NLTT5306. \textbf{Top Right:} Flux contributions from the day and night hemispheres as a function of inclination. As we increase inclination, we see more of the day side and less of the night side. \textbf{Bottom Right:} Flux correction factors for each non-irradiated brown dwarf temperature considered in the left panel, determined by the expression in the upper right corner. Both the low ($i=58^{\circ}$) and high ($i=79.7^{\circ}$) ends of the possible inclination range for NLTT5306 are shown as diamonds and stars, respectively. From this panel, we conclude that the flux correction factor for NLTT5306B could be between 7 and 25\%.}
\label{fig:effect_inc}
\end{center}
\end{figure}

Exploring the effects of inclination on observational constraints and results provides a more accurate view of systems and their dynamics. To get a better understanding of how unknown inclination impacts our results, we created an exploratory atmosphere model. In this basic model, there are only two hemispheres, day and night, where the night-side temperature is determined by the internal brown dwarf temperature (i.e. non-irradiated temperature) and the day-side is exposed to the same amount of external irradiation as NLTT5306B. In the absence of atmospheric circulation, the irradiation temperature at the substellar point on NLTT5306B is around 910~K\footnote{Substellar irradiation temperature based on orbital separation $a = 0.566$ R$_{\odot}$, white dwarf temperature $T_{\rm{WD}} = 7756$ K, and radii of both dwarfs: $R_{\rm{BD}} = 0.095$ R$_{\odot}$ and $R_{\rm{WD}} = 0.016$ R$_{\odot}$.}. Depending on the internal brown dwarf temperature, 910~K could create a moderate to large flux contrast between the day- and night-sides. We show this effect for range of non-irradiated brown dwarf temperatures from 900 to 1600 K, coupled with the effect of inclination in the left panel of Figure~\ref{fig:effect_inc}. 

The fluxes for day and night, F$_{\rm{Day}}$ and F$_{\rm{Night}}$, were calculated by creating a basic version of the atmosphere model from Section~\ref{sec:longtempdist} for each non-irradiated BD temperature. This version of the atmosphere model ignores $A_B$ and $f_{\rm{irr-red}}$. We then took the hemisphere-integrated temperature for both day- and night-sides and converted them into fluxes via F$\propto$T$^4$. We imagined a system orientation where the the brown dwarf is farthest from the observer. Figure~\ref{fig:effect_inc} shows the total observed flux as a function of inclination, according to the following equations: 
\begin{equation}
    \begin{aligned}
        F(i) &= F_{\rm{Day}}(i) + F_{\rm{Night}}(i),\\
        F_{\rm{Day}}(i) &= F_{\rm{Day}}(90^{\circ})\Big(1 - \frac{1}{2}\rm{cos}(i)\Big),\\
        F_{\rm{Night}}(i) &= F_{\rm{Night}}(90^{\circ})\Big(\frac{1}{2}\rm{cos}(i)\Big).
    \end{aligned}
\end{equation}
For equatorial viewing geometry ($i=90^\circ$), only the day-side would be visible. However, for any other inclinations, a fraction of the night-side hemisphere would be simultaneously visible, causing the observed hemisphere to appear dimmer than it really is (illustrated in the top right panel of Figure~\ref{fig:effect_inc}). In contrast, if we were to observe the night-side hemisphere at inclinations other than 90$^{\circ}$, the partially visible day-side would cause the hemisphere to appear brighter. Our model accounts for this and helps to assess the importance of this effect. 

Based on our exploratory model, we find that the effect of uncertain inclination (within a constrained range) would impact the derived fluxes by 7 to 25\% (relative to the equatorial view), as shown in the bottom right panel of Figure~\ref{fig:effect_inc}. Thus, to obtain a more physically accurate view of NLTT5306B's atmosphere, we included the inclination as a parameter in the simple atmosphere model. Through a grid search, we constrained possible values for the Bond albedo ($A_B$), irradiation redistribution fraction ($f_{\rm{irr-red}}$), non-irradiated BD temperature ($T_{\rm{non-irr}}$), and inclination ($i$). Inclination was constrained to $78.0^{+1.1}_{-7.6}$ degrees (see \ref{sec:longtempdist}). At $78.0^{\circ}$ inclination, our observed day- and night-side flux values would need to be corrected by $+10\%$ (for day-side) and $-10\%$ (for night-side) in order to get the likely true flux for each hemisphere.

\begin{figure}
\begin{center}
\includegraphics[width=0.95\textwidth]{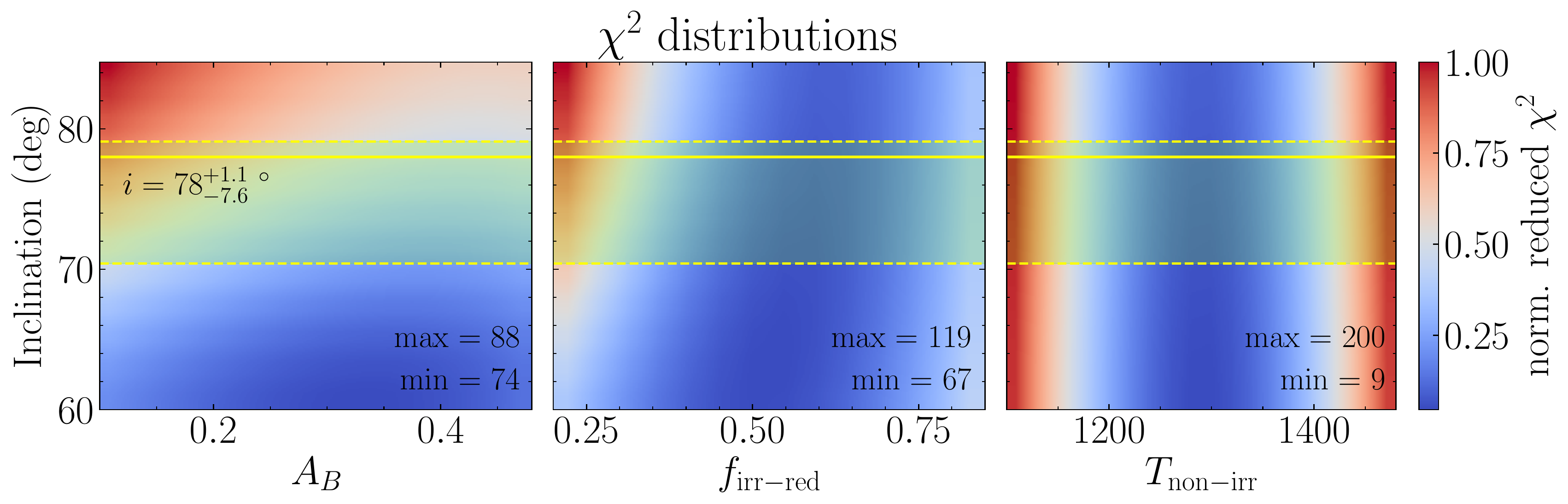}
\caption{Distribution of normalized chi-squared values from the atmospheric model grid search conducted in Section~\ref{sec:longtempdist}. Minimum and maximum values used to normalize each distribution are labeled in the bottom right of each panel. Panels show relationships between inclination and the other three parameters: Bond albedo ($A_B$), fraction of irradiation redistributed from day-to-night ($f_{\rm{irr-red}}$), and non-irradiated BD temperature ($T_{\rm{non-irr}}$). Lower values of chi-squared indicate better fits to the observations. Best-fit inclination for NLTT5306B is shown as a solid yellow line, with 1-sigma uncertainty shown as dashed yellow lines. In the middle and right panels, there is little to no correlation between these parameters and inclination. However, the left-most panel shows that Bond albedo, $A_B$, is mildly sensitive to the inclination.}
\label{fig:inclination_covariance}
\end{center}
\end{figure}

Next, after we ran the grid model search with all 4 parameters (see Section~\ref{sec:longtempdist}), we determined how correlated the inclination is with the other parameters. We searched for correlation by using the four-dimensional grid of chi-squared values, collapsed down to the 2 axes of interest. The result of this method is shown in Figure~\ref{fig:inclination_covariance}. We used dashed lines to indicate the likely range of inclinations. We find little to no correlation between $i$ inclination and the following two parameters: irradiation redistribution fraction ($f_{\rm{irr-red}}$) and non-irradiated BD temperature ($T_{\rm{non-irr}}$). Therefore, the derived irradiation redistribution fraction and non-irradiated BD temperature are insensitive to the uncertainty in the inclination and are thus robust values. However, the derived Bond albedo is somewhat sensitive to the exact inclination, with the $A_B$ varying between 0.15 to 0.47 for likely inclination range of $78.0^{+1.1}_{-7.6}$. 

\bibliography{biblio}
\bibliographystyle{aasjournal}

\end{document}